\documentclass[pra,aps,10pt,twocolumn,floatfix,superscriptaddress,
notitlepage,longbibliography]{revtex4-2}

\usepackage{xcolor, soul}

\usepackage[T1]{fontenc}
\usepackage[utf8]{inputenc}
\usepackage{graphicx,color}
\usepackage{amsmath,amssymb,amsfonts}
\usepackage{quantikz}
\usepackage{amsthm}
\usepackage{subcaption}
\newtheorem{definition}{Definition}
\usepackage{bm}
\usepackage[linesnumbered,ruled,vlined]{algorithm2e}
\usepackage[noend]{algpseudocode}
\usepackage[colorlinks=true,citecolor=blue,linkcolor=magenta,urlcolor=blue]{hyperref}
\usepackage{braket}
\usepackage{tikz}
\usepackage{booktabs}
\usepackage{multirow}
\usepackage{siunitx}
\usepackage{eurosym}

\begin{document}


\title{Shallow instantaneous quantum polynomial-time circuits for generative modeling on noisy intermediate-scale quantum hardware
}

\author{Oriol Balló-Gimbernat}
\email{oriol.ballo@eurecat.org}
\affiliation{Eurecat, Centre Tecnològic de Catalunya, Barcelona, Spain}
\affiliation{Centre de Visió per Computador (CVC), Barcelona, Spain}
\affiliation{Universitat Autònoma de Barcelona (UAB), Barcelona, Spain}

\author{Marcos Arroyo-Sánchez}
\affiliation{Eurecat, Centre Tecnològic de Catalunya, Barcelona, Spain}
\affiliation{Centre de Visió per Computador (CVC), Barcelona, Spain}
\affiliation{Universitat Autònoma de Barcelona (UAB), Barcelona, Spain}

\author{Paula García-Molina}
\affiliation{Centre de Visió per Computador (CVC), Barcelona, Spain}

\author{Adan Garriga}
\affiliation{Eurecat, Centre Tecnològic de Catalunya, Barcelona, Spain}

\author{Fernando Vilariño}
\affiliation{Centre de Visió per Computador (CVC), Barcelona, Spain}
\affiliation{Universitat Autònoma de Barcelona (UAB), Barcelona, Spain}

\date{\today}

\begin{abstract}

Generative modeling is one of the most promising applications of quantum machine learning, yet training and deploying Quantum Generative Models (QGMs) on near-term hardware remains effectively intractable due to prohibitive gradient estimation and implementation costs. We propose a resource-efficient approach based on shallow Instantaneous Quantum Polynomial-time (IQP) circuits that circumvents these bottlenecks by leveraging efficient classical training while retaining the guarantee of sampling hardness. To validate this approach, we formalize graph generation as a hierarchy of physical correlations, allowing us to map abstract data features---such as edge density and bipartiteness---directly to the quantum observables required to learn them. We validate our protocol through demonstrations both on real hardware (from $28$ to $153$ qubits) and simulations ($28$ qubits). Results show that while global structural features exhibit significant degradation beyond $91$ qubits, our models achieve high-precision reproduction of local correlations, even up to $153$ qubits. These findings establish shallow IQP circuits as a robust, scalable candidate for generative tasks on current Noisy Intermediate-Scale Quantum (NISQ) devices.
\end{abstract}

\maketitle
\section{Introduction}

\begin{figure*}[ht!]
    \centering
    \includegraphics[width=.95\linewidth]{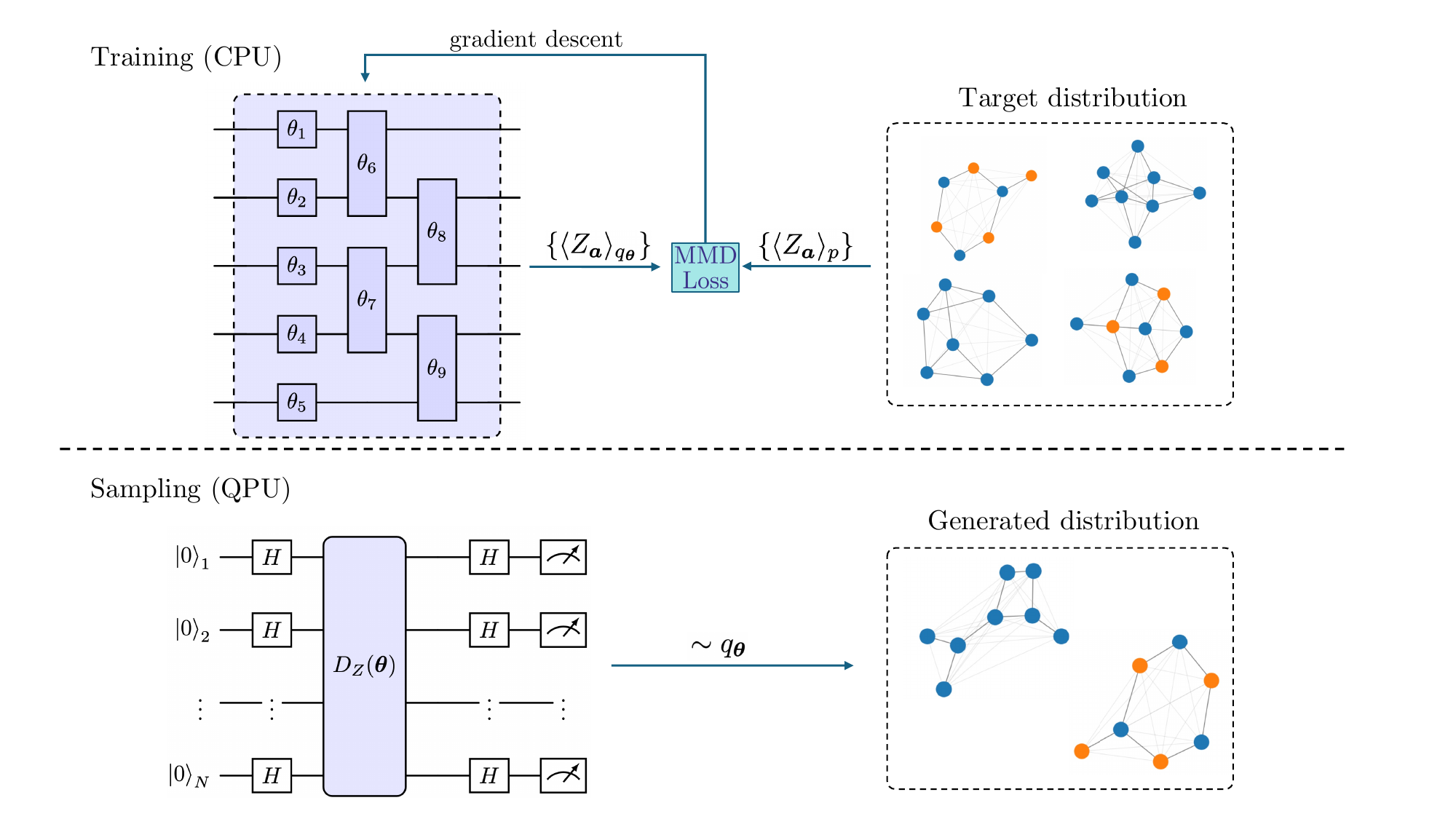}
    \caption{Overview of the proposed workflow. \textbf{Top}: Model training is performed entirely on classical hardware via a classical estimation of the MMD loss. \textbf{Bottom}: Sampling is carried out on IBM’s Aachen quantum processor.}
    \label{fig:workflow}
\end{figure*}

Generative modeling stands as a cornerstone of unsupervised machine learning, enabling the synthesis of data that captures the underlying statistics of complex distributions. Within this domain, Quantum Generative Models (QGMs) have emerged as a promising frontier, leveraging quantum computation to represent probability distributions that are potentially intractable for classical systems \cite{zoufalGenerativeQuantumMachine2021, swekeQuantumClassicalLearnability2021, zengLearningInferenceGenerative2019}. A particularly compelling architecture in this landscape is the Quantum Circuit Born Machine (QCBM) \cite{coyleBornSupremacyQuantum2020, mitaraiQuantumCircuitLearning2018a, liuDifferentiableLearningQuantum2018a, kissConditionalBornMachine2022}. However, training QCBMs on near-term hardware presents significant hurdles, primarily due to the poor scaling of gradient estimation costs and the prevalence of barren plateaus, which currently restricts training to small-scale models. Furthermore, even in instances where large-scale models can be trained without barren plateaus, efficient implementation and robust guarantees of classical hardness remain elusive \cite{cerezoDoesProvableAbsence2025}.

In this work, we propose a novel generative modeling approach that overcomes these bottlenecks by leveraging the unique properties of Instantaneous Quantum Polynomial-time (IQP) circuits. IQP circuits occupy a prime position for near-term applications: they are conjectured to be hard to sample from classically \cite{shepherdInstantaneousQuantumComputation2009, marshallImprovedSeparationQuantum2024}, yet their specific algebraic structure permits the efficient classical calculation of certain output statistics that can be used for training \cite{recio-armengolTrainClassicalDeploy2025, kastureProtocolsClassicallyTraining2023}. We introduce a hybrid workflow based on shallow IQP circuits that exploits this duality---training the model on classical hardware and deploying it on quantum processors for sampling \cite{recio-armengolTrainClassicalDeploy2025, recio-armengolIQPoptFastOptimization2025}---while minimizing the resources required for implementation and retaining classical hardness. This approach effectively circumvents the scalability and implementation challenges inherent in standard QCBMs.

To rigorously validate this approach and characterize the expressivity of shallow IQP circuits, we use graph generation as a testbed. Graphs represent relationships between entities and provide a unified framework to describe complex systems via their components and interactions \cite{newmanNetworks2018}. Their versatility has driven the development of graph generation methods for applications ranging from drug discovery to scheduling \cite{drobyshevskiyRandomGraphModeling2020, bonifatiGraphGeneratorsState2021, bonginiMolecularGenerativeGraph2021, yangMoleculeGenerationDrug2024, cordeiroRandomGraphGeneration2010}. Learning a graph distribution remains challenging because node interactions span both local and global dependencies \cite{groverGraphiteIterativeGenerative2019, tranDeepNCDeepGenerative2020, bacciuEdgebasedSequentialGraph2020, simonovskyGraphVAEGenerationSmall2018, flam-shepherdGraphDeconvolutionalGeneration2020, niuPermutationInvariantGraph2020}. As the number of nodes increases, the combinatorial complexity of pairwise and higher-order relationships grows rapidly: adding a single node to an $M$-node graph introduces $M$ new potential edges, so that the number of possible configurations increases from $2^{\binom{M}{2}}$ to $2^{\binom{M}{2}} \cdot 2^M$. This exponential scaling implies that even modest increases in graph size can dramatically expand the hypothesis space. Although practical problems typically restrict attention to structured graph families, models must still possess sufficient expressivity to navigate this high-dimensional space \cite{guoSystematicSurveyDeep2022, simonovskyGraphVAEGenerationSmall2018}.

Rather than treating graph generation  as a downstream application, we utilize it to map abstract data features to physical correlations. Simple properties like edge density correspond to local ($1$-body) correlations, while structural constraints like bipartiteness correspond to global ones. This formalization allows us to disentangle the model's capacity to learn features of different complexities.

We demonstrate the efficacy of this approach by training shallow IQP models to generate random bipartite and Erd\H{o}s–Rényi graph distributions. Our demonstration results, obtained on an IBM Heron r2 $156$ qubit processor, show that the proposed models are highly effective at reproducing low-bodied correlations, even up to $153$ qubits. While global structural features exhibit higher performance degradation beyond $91$ qubits, due to hardware noise. These findings establish shallow IQP circuits as a resource-efficient candidate for near-term generative tasks.

The paper is organized as follows. Section~\ref{sec:graph-dist} defines the target graph distributions, their representations, and our features of interest. Section~\ref{sec:iqp-qcbms} describes the proposed shallow IQP model and its optimization. Section~\ref{sec:experimental-setup} outlines the demonstration setup, including the datasets, training, hyperparameter optimization, and sampling. Section~\ref{sec:results} reports the results. Finally, section~\ref{sec:conclusion} concludes and discusses future directions.

\section{Graphs and Random Graph Distributions}
\label{sec:graph-dist}

\begin{figure*}[ht!]
    \centering
    \includegraphics[width=.7\linewidth]{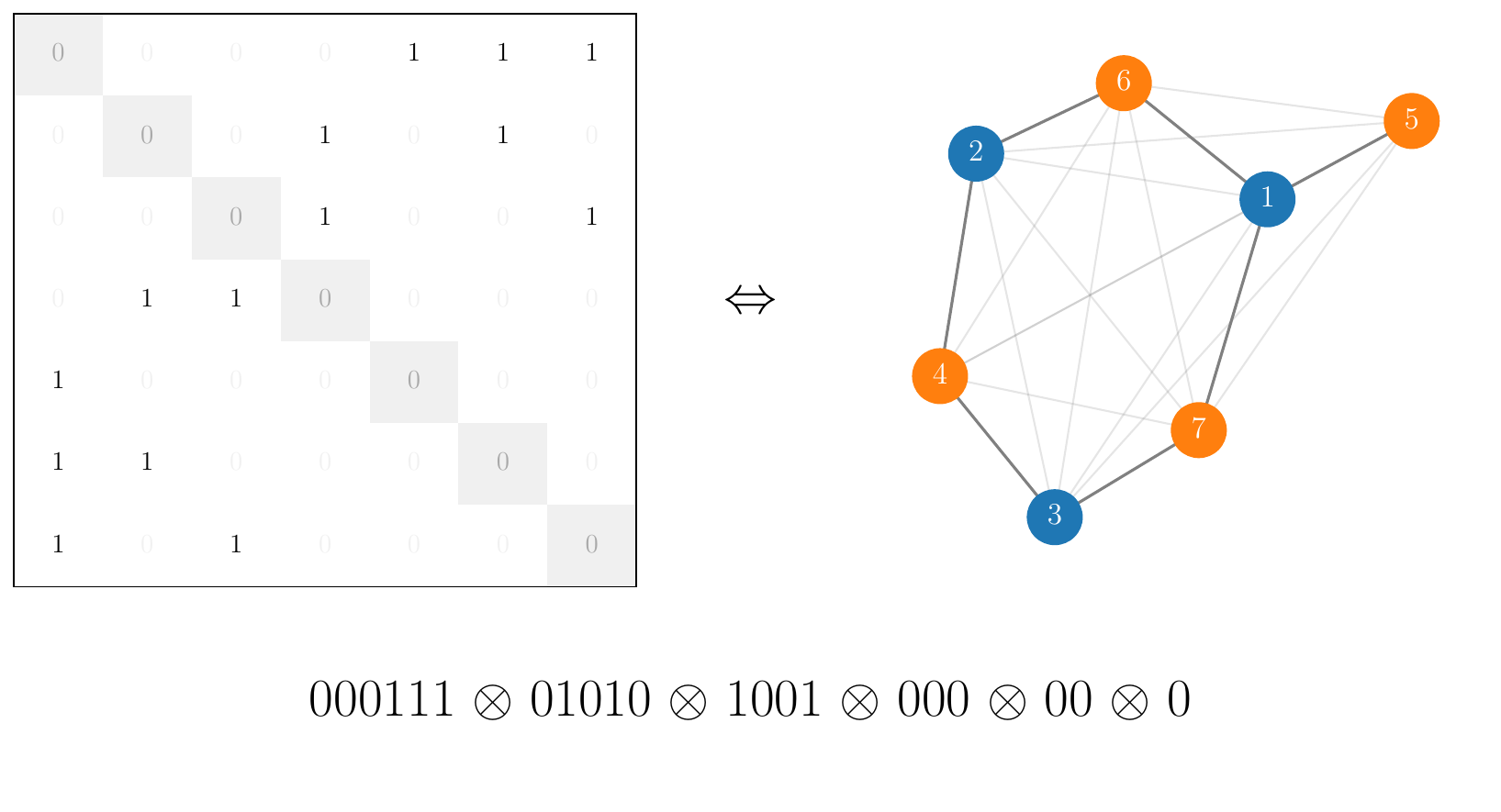}
    \caption{\textbf{Top}: Adjacency matrix representation of a $7$-node undirected graph. \textbf{Bottom}: Corresponding bit string encoding constructed from the upper-triangular part of the adjacency matrix.}
    \label{fig:graph-equivalence}
\end{figure*}

A graph distribution defines a probability distribution over the space of graphs satisfying a given criterion. In this work, we restrict attention to undirected, unweighted graphs to isolate structural dependencies while maintaining a compact binary representation. Such graphs connect nodes by undirected edges: if node $i$ is linked to node $j$, then node $j$ is equally linked to node $i$. Consequently, the adjacency matrix $\mathbf{A}$ is symmetric, $A_{ij} = A_{ji}$, and self-loops are excluded, i.e. $A_{ii} = 0\  \forall\ i$.

Because of this symmetry, specifying one triangular portion of $\mathbf{A}$ suffices. An undirected graph with $M$ nodes is therefore fully characterized by
\begin{equation}
    N = \frac{M(M-1)}{2},
\end{equation}
binary variables that encode the presence or absence of edges between distinct nodes (see Fig.~\ref{fig:graph-equivalence}). This encoding defines the effective dimensionality of the generative problem.

We focus on two families of random graph distributions. The first is the Erd\H{o}s–Rényi (ER) model \cite{erdosRandomGraphs2022}, which defines graphs in which each possible edge is included independently with probability $\rho \in [0, 1]$, known as the edge density. The second family comprises bipartite (BP) graphs, whose vertices can be partitioned into two disjoint sets such that edges exist only between and never within---equivalently, 2-colorable graphs. BP distributions impose structural constraints that contrast with the edge-independent probabilistic structure of ER graphs, making them a useful testbed.

These two classes are not mutually exclusive: an ER sample may be bipartite, particularly at low edge densities, while any BP graph can be viewed as a constrained instance of the ER model. Throughout this work, BP distributions refer to ensembles of random bipartite graphs, whereas ER distributions denote ensembles of random graphs that may or may not exhibit bipartite structure.

\subsection{Bit string graph representation}
\label{sec:graph-rep}
To map the circuit output to a graph, we flatten the upper triangular part of the adjacency matrix row by row into a binary string $\mathbf{z} = (z_1, \dots, z_N)$, where $N = M(M-1)/2$ and $M$ is the number of nodes, as illustrated in Fig.~\ref{fig:graph-equivalence}. In this representation, each qubit in the quantum circuit corresponds to a potential edge, and a measurement in the computational basis yields a binary string that uniquely defines a specific graph.

This encoding scales quadratically with the number of nodes. While more compact latent encodings or reparameterizations could reduce the required qubit count, we adopt this direct mapping as an interpretable baseline. It ensures an unambiguous correspondence between quantum measurement outcomes and graph structures, avoiding potential confounding factors introduced by latent representations.

\subsection{Features of interest}
Graph features represent different degrees of node interactions: some depend only on local edge probabilities, while others require knowledge of global relationships. 

We formalize this notion as follows. Let $P(\mathbf{z})$ denote the target distribution over adjacency vectors, $\mathbf{z} = (z_1, \ldots, z_N) \in \{0,1\}^N$, 
and let $f : \{0,1\}^N \to \mathbb{R}$ be a feature of interest.

\begin{definition}[$k$-bodied feature]
A feature $f$ is $k$-bodied if its expected value under $P$, $\mathbb{E}_{P}[f(\mathbf{z})]$, depends only on correlations among at most $k$ edges.

Formally, for any subset of edge indices $S \subseteq \{1, \ldots, N\}$ with $|S| \leq k$, let $P_S(\mathbf{z}_S)$ denote the marginal distribution of the subvector $\mathbf{z}_S = (z_i : i \in S)$. Then $f$ is $k$-bodied if the collection of all such marginals,
\begin{equation}
    \big\{ P_S(\mathbf{z}_S) : S \subseteq \{1, \ldots, N\},\, |S| \le k \big\},
\end{equation}
is sufficient to uniquely determine $\mathbb{E}_{P}[f(\mathbf{z})]$.
\end{definition}

\begin{definition}[Global feature]
A feature $f$ is said to be global if it is not $k$-bodied for any fixed $k < N$. 
\end{definition}

These definitions measure the correlations required to learn to reproduce any given property in an idealized setting. We now describe the features used in this work, their theoretical bodyness, and practical considerations.

The simplest feature is density, $\rho$, which measures the fraction of present edges relative to the maximum number of edges,
\begin{equation}
    \rho(\mathbf{z}) = \frac{2}{M(M-1)} \sum_{i=1}^N z_i.
\end{equation}
Density provides a coarse summary of the overall connectivity, showing whether a graph is sparse ($\rho\approx 0$) or dense ($\rho\approx 1$), but it does not reveal how edges are arranged among nodes. Since the expected density depends only on single-edge probabilities, it is a $1$-bodied feature. Consequently, it should be easy to learn and reproduce.

The next feature of interest is the degree distribution, which captures the number of edges incident on each node and provides a more detailed view of connectivity than density alone. For a node $v_j$, the degree is defined as 
\begin{equation}
    \deg(v_j)= \sum_{i\in\mathcal{E}_j}z_i, 
\end{equation}
where $\mathcal{E}_j$ is the set of edges incident on $v_j$. The degree distribution of a graph is then the histogram of
$\deg(v_j)$ over all nodes $j\in\{1,\dots,M\}$. For random graphs with independent edges, the degree of a node is the sum of independent Bernoulli variables, yielding a binomial distribution. Therefore, the degree distribution is also $1$-bodied. Although it has the same bodyness as the density, it should be harder to reproduce in practice, as it encodes additional information beyond the mean edge probability. 

Another feature we consider is the bipartite property, which indicates whether a graph is $2$-colorable---i.e., has no odd-length cycles. The degree of correlations required to determine this property increases with graph size; for an $M$-node graph, one must capture all odd-cycle correlations up to length $M$, or $M-1$ if $M$ is even. Consequently, bipartiteness is a global feature, as it depends on correlations that span the entire graph. 

Among the considered features, bipartiteness is the most difficult to reproduce. It's a binary property---a graph is either bipartite or not---plus it is highly sensitive to perturbations: even a single bit flip can introduce an odd-length cycle, converting a bipartite graph into a non-bipartite one.

\section{IQP circuit born machines}
\label{sec:iqp-qcbms}

A QCBM is a quantum generative model that represents a probability distribution in the measurement statistics of a pure quantum state \cite{liuDifferentiableLearningQuantum2018}. Sampling from this distribution is achieved through projective measurements in the computational basis, producing single-shot samples. 

In this work, we employ QCBMs constructed with IQP circuits. These represent a class of quantum computations that are restricted to commuting operations \cite{shepherdInstantaneousQuantumComputation2009, nakataDiagonalQuantumCircuits2014}. Despite this apparent limitation, they have emerged as a powerful tool for generative modeling. Sampling from most instances of IQP circuits is conjectured to be classically hard \cite{bremnerAverageCaseComplexityApproximate2016}, as simulating them efficiently would imply a collapse of the polynomial hierarchy to its second level \cite{bremnerClassicalSimulationCommuting2011, marshallImprovedSeparationQuantum2024}. Yet, classical computers can efficiently estimate the expectation values of their commuting operators \cite{nestSimulatingQuantumComputers2010}. This feature enables their classical optimization when the cost function depends solely on those values \cite{recio-armengolIQPoptFastOptimization2025}. Thus, it allows the implementation of a hybrid framework where classical resources are used for training and quantum resources for sampling.

Our objective is to train shallow IQP models to generate graphs according to a ground truth distribution $p$. Specifically, we aim to learn a parameterized distribution $q_{\boldsymbol{\theta}}$ such that, for any feature of interest $f(\mathbf{z})$,
 
\begin{equation}
 \left|\mathbb E_{\mathbf z\sim q_{\boldsymbol{\theta}}}[f(\mathbf z)] - \mathbb E_{\mathbf z\sim p}[f(\mathbf z)]\right| < \varepsilon,
\end{equation} 
where $\varepsilon$ defines the target accuracy. 

\subsection{Shallow IQP circuits}
The foundation of our models is a $N$-qubit shallow IQP circuit, building on the general definition from Nakata \textit{et al.} \cite{nakataDiagonalQuantumCircuits2014}.

\begin{definition}
A shallow IQP circuit on $N$ qubits is defined by the following sequential operations,
\begin{enumerate}
    \item \textbf{Basis Transformation:} An initial layer of Hadamard gates $H^{\otimes N}$ applied to all qubits.
    

    \item {\textbf{Diagonal Evolution:} A parameterized unitary block $D_Z(\boldsymbol{\theta})$ diagonal in the computational basis. This block consists of a sequence of commuting rotation gates generated by Pauli-$Z$ strings acting on subsets of qubits $S$:
        \begin{equation}
            D_Z(\boldsymbol{\theta}) = \prod_{S \in \mathcal{I}} \exp\left(-i \theta_S \prod_{j \in S} Z_j\right),
        \end{equation}
        where $\mathcal{I}$ is a collection of subsets of $\{1, \dots, N\}$ specifying the interactions. The collection $\mathcal{I}$ is chosen such that the total circuit depth is $\mathcal{O}(1)$, independent of the number of qubits $N$.}
    
    \item \textbf{Inverse Transformation:} A final layer of Hadamard gates $H^{\otimes N}$ applied to all qubits.
    
    \item \textbf{Measurement:} A projective measurement in the computational basis.
\end{enumerate}
\end{definition}

\begin{figure}[h]
    \centering
    \includegraphics[width=0.7\linewidth]{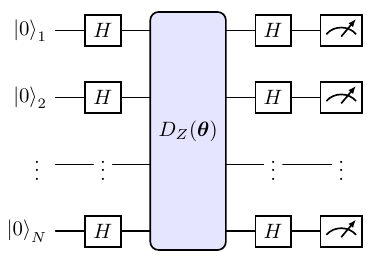}
    \caption{Circuit diagram of a parameterized IQP circuit. The unitary $D_Z(\boldsymbol{\theta})$ contains all diagonal parameterized gates. For constant-depth $D_Z(\boldsymbol{\theta})$, the circuit realizes a shallow instance of the IQP model.}
    \label{fig:iqp-diagram}
\end{figure}

The circuit diagram form is illustrated in Figure \ref{fig:iqp-diagram}.

This sequence of operations implements a unitary $U(\boldsymbol{\theta})=H^{\otimes N}D_Z(\boldsymbol{\theta})H^{\otimes N}$, where $\boldsymbol \theta = \{\theta_i\}_{i=1}^n$, that prepares the following quantum state,
\begin{equation} \label{eq:iqp_state}
    |\Psi(\boldsymbol{\theta})\rangle = H^{\otimes N} D_Z(\boldsymbol{\theta}) H^{\otimes N} |0\rangle^{\otimes N},
\end{equation}
which encodes a probability distribution with respect to measurements in the computational basis.

Crucially, for any observable $O_z$ that is a product of Pauli-Z operators, its expectation value,
\begin{equation}
    \langle \Psi(\boldsymbol\theta)|O_z|\Psi(\boldsymbol\theta)\rangle,
\end{equation}
can be computed efficiently on classical hardware. The precise complexity of this process is detailed in Proposition 2 of Recio-Armengol \textit{et al.} \cite{recio-armengolIQPoptFastOptimization2025}.

All parameterized operations are contained within the diagonal gate block. Therefore, the ansatz is fully specified by this block alone. Crucially, it determines both the expressivity and the hardware efficiency of a parameterized circuit. A simple ansatz may be easy to implement but fail to capture high-bodied correlations, whereas a more complex one may be able to represent such correlations but at the cost of increased resource requirements---and an increased noise sensitivity. To balance this trade-off, we employ a shallow ansatz that uses few resources while still being conjectured to be classically intractable.

{ The classical simulability of IQP circuits depends strictly on the connectivity and type of gates employed, rather than circuit depth, as the commuting nature of the gates renders the temporal order of operations irrelevant. Circuits composed solely of single-qubit rotations are trivially simulable, as they generate no entanglement. Furthermore, circuits restricted to nearest-neighbor two-qubit gates on a planar grid can also be efficiently simulated, regardless of the number of layers. This is because their output probability distribution maps to the partition function of a 2D Ising model with zero magnetic field, a problem exactly solvable via the Pfaffian method (free fermions) \cite{fujiiQuantumCommutingCircuits2017}.

Notably, combining these two simulable designs in a planar 2D layout pushes the model out of the simulable regime \cite{fujiiQuantumCommutingCircuits2017}. Introducing single-qubit rotations corresponds to adding magnetic fields to the underlying Ising model. This modification breaks the free-fermion mapping and renders the calculation of the partition function \#P-hard \cite{fujiiQuantumCommutingCircuits2017}. This combination—nearest-neighbor interactions plus single-qubit rotations—constitutes the design chosen for this work, ensuring the ansatz remains shallow yet conjectured to be classically intractable.}

This ansatz maps efficiently to the target quantum processor (QPU), IBM's Aachen, since both parameterized gates belong to the device's native gate set. In particular, RZ gates are implemented as virtual rotations with negligible latency and error \cite{mckayEfficientZGatesQuantum2017}. Although Hadamard gates are not native, the compiler decomposes them into RZ($\pi/2$) and SX gates, introducing only a minimal depth overhead.

Notably, we omit ancilla qubits, even though they are necessary to achieve universality \cite{kurkinNoteUniversalityParameterized2025}. While this choice limits expressive power, it may also enhance trainability---as expressivity and barren plateaus are closely linked \cite{laroccaBarrenPlateausVariational2025, cerezoDoesProvableAbsence2025}. 

\subsection{Optimization via the maximum mean discrepancy}
\label{sec:mmd}

To train an IQP model on classical hardware, one must formulate the optimization task in terms of expectation values---both from the target data and the circuit itself.

The maximum mean discrepancy (MMD) can be formulated in such terms. It is an integral probability metric that compares two probability distributions, $p$ and $q$, based on samples, $\boldsymbol x$ and $\boldsymbol y$, drawn from each in a reproducing kernel Hilbert space (RKHS) \cite{JMLR:v13:gretton12a}. The squared MMD is defined as \cite{sutherlandGenerativeModelsModel2021}
\begin{equation}
\begin{split}
    \text{MMD}^2(p,q_\theta) &= \mathbb E_{\boldsymbol{x}\sim p,\boldsymbol y\sim p}[k(\boldsymbol x,\boldsymbol y)] \\
               &\quad - 2\mathbb E_{\boldsymbol x\sim p, \boldsymbol y\sim q_\theta}[k(\boldsymbol x,\boldsymbol y)] \\
               &\quad + \mathbb E_{\boldsymbol x\sim q_\theta, \boldsymbol y\sim q_\theta}[k(\boldsymbol x,\boldsymbol y)],
\end{split}
\end{equation}
where $k(\boldsymbol x, \boldsymbol y)$ is a positive definite kernel. If $k$ is characteristic, then $\text{MMD}^2=0$ if and only if $p=q_\theta$ \cite{JMLR:v13:gretton12a}. The three terms respectively measure intra-data similarity, model-data cross-similarity, and intra-model similarity.

Central to this approach, we use the Gaussian kernel, 
\begin{equation}
    k_\sigma(\boldsymbol x, \boldsymbol y) = \exp{-\frac{||\boldsymbol x-\boldsymbol y||^2}{2\sigma^2}},
\end{equation}
for vectors $\boldsymbol x, \boldsymbol y \in \{0,1\}^n$. Since each bit contributes either $0$ or $1$ to the sum, the squared Euclidean distance $\|\boldsymbol x - \boldsymbol y\|^2$ coincides with the Hamming distance, i.e., the number of positions at which the entries of the vectors differ. This equivalence allows the MMD to be expressed in terms of Pauli-Z observables \cite{rudolphTrainabilityBarriersOpportunities2024}, 
\begin{equation}
\label{eq:mmd_iqp}
\text{MMD}^2(\boldsymbol\theta) = \mathbb{E}_{\boldsymbol a\sim\mathcal{P}{\sigma}(\boldsymbol a)}\left[(\langle Z_{\boldsymbol a} \rangle_p - \langle Z_{\boldsymbol a} \rangle_{q_{\boldsymbol \theta}})^2\right],
\end{equation}
where $\boldsymbol{a}\in\{0,1\}^n$ encodes a `mask' that corresponds to a Pauli-Z observable,
\begin{equation}
    Z_{\boldsymbol{a}}= \prod_{i=1}^nZ_i^{a_i}.
\end{equation}

The expectation value measures parity correlations among the selected bits,
\begin{equation}
    \langle Z_{\boldsymbol a}\rangle_p=\mathbb E_{\boldsymbol x\sim p}[(-1)^{\boldsymbol a \cdot \boldsymbol x}],
\end{equation}
where $(-1)^{\boldsymbol{a} \cdot \boldsymbol{x}} = +1$ for even parity and $-1$ for odd parity.

Crucially, the weighting distribution $\mathcal P_\sigma$ determines which correlations the loss emphasizes,
\begin{equation}
    \mathcal P_{\sigma}(\boldsymbol{a})=(1-p_\sigma)^{n-|\boldsymbol{a}|}(p_\sigma)^{|\boldsymbol{a}|}, \quad p_\sigma= \frac{1-  e^{-1/2\sigma^2}}{2}.
\end{equation}
Each bit appears independently with probability $p_\sigma$, so the active bit count $|\boldsymbol{a}|$ follows a binomial distribution with mean $np_\sigma$ \cite{rudolphTrainabilityBarriersOpportunities2024}. { A derivation of the MMD as a quantum observable is shown in the Appendix} \ref{appendix_mmd}.

This behavior has a profound impact on training. When $\sigma$ is constant, $\sigma \in O(1)$, the MMD emphasizes high-order correlations, effectively making the loss global \cite{rudolphTrainabilityBarriersOpportunities2024}. Such cost function is known to yield barren plateaus in unstructured circuits, potentially rendering the model untrainable \cite{cerezoCostFunctionDependent2021}. Conversely, if $\sigma$ scales with the number of qubits, $\sigma \in O(n)$, the MMD primarily captures low-order correlations, ensuring non-vanishing gradients and preserving trainability. However, it blinds the loss to the higher-order correlations.

This observation directly connects to the learnability of the target properties. Features such as density are expected to be learnable in a regime free of barren plateaus, whereas bipartiteness may require a constant $\sigma$ and thus lie in an untrainable regime.

\begin{table*}[hbt!]
  \centering
  \setlength{\tabcolsep}{3pt}
  \renewcommand{\arraystretch}{1.1}
    \caption{Summary of the datasets. $\bar\rho$ denotes the average density of the samples, BP (\%) the fraction of bipartite graphs, $\bar\beta$ the average spectral bipartivity, and $N$ the total number of samples.}
  \label{tab:dataset-info}
  \begin{tabular}{c c *{3}{ S c c c }}
    \toprule
    Nodes & Type
      & \multicolumn{4}{c}{Dense}
      & \multicolumn{4}{c}{Medium}
      & \multicolumn{4}{c}{Sparse} \\
    \cmidrule(lr){3-6}
    \cmidrule(lr){7-10}
    \cmidrule(lr){11-14}
    & 
      & {$\bar\rho$}
      & {BP.\%}
      & {$\bar\beta$}
      & {N}
      & {$\bar\rho$}
      & {BP.\%}
      & {$\bar\beta$}
      & {N}
      & {$\bar\rho$}
      & {BP.\%}
      & {$\bar\beta$}
      & {N} \\
    \midrule
    \multirow{2}{*}{8}
      & BP
        & 0.3110 & 100 &   1.0 &   261
        & 0.2887 & 100 &   1.0 &   271
        & 0.2256 & 100 &   1.0 &   133\\
      & ER
        & 0.7618 & 0.0 &    0.55 &   200
        & 0.4420 & 0.5 &    0.79 &   200
        & 0.2207 & 51.5 &   0.95 &   200 \\
    \midrule
    \multirow{2}{*}{10}
      & BP
        & 0.3470 & 100 & 1.0 & 498
        & 0.2307 & 100 & 1.0 & 500
        & 0.1771 & 100 & 1.0 & 473 \\
      & ER
        & 0.7884 & 0.0 & 0.51 & 500
        & 0.4380 & 0.2 & 0.69 & 500
        & 0.1919 & 33.8 & 0.94 & 500 \\
    \midrule
    \multirow{2}{*}{14}
      & BP
        & 0.3727 & 100 & 1.0 & 995
        & 0.2084 & 100 & 1.0 & 999
        & 0.1042 & 100 & 1.0 & 995 \\
      & ER
        & 0.8145 & 0.0 & 0.50 & 1000
        & 0.4487 & 0.0 & 0.57 &   1000
        & 0.1575 & 27.9 & 0.93 &   1000\\
    \midrule
    \multirow{2}{*}{18}
      & BP
        & 0.3697 & 100 & 1.0 & 995
        & 0.1967 & 100 & 1.0 & 998
        & 0.0747 & 100 & 1.0 & 992\\
      & ER
        & 0.8255 & 0.0 & 0.5 &   1000
        & 0.4428 & 0.0 & 0.53 &   1000
        & 0.1497 & 17.7 & 0.89 &   1000\\
    \bottomrule
  \end{tabular}
\end{table*}

\section{Demonstration setup}
\label{sec:experimental-setup}

This section describes the demonstration framework used to evaluate the proposed workflow (Fig.~\ref{fig:workflow}). We begin by outlining the generation of synthetic graph datasets employed for training and validation. The following subsections detail the model training procedure, hyperparameter optimization (HPO), and sampling strategy. We then introduce the quantitative metrics used to compare the generated and target graph distributions. Together, these components form a reproducible pipeline for assessing the model’s performance.

All the code and data supporting this work are openly available in Zenodo \cite{ballo_gimbernat_2026_zenodo}.

\subsection{Datasets}

We generated synthetic datasets for the two graph families introduced in Sec.~\ref{sec:graph-dist}: Erd\H{o}s–Rényi (ER) graphs, where each edge is included independently with probability $\rho$, and bipartite (BP) graphs, where a $2$-colorability constraint is imposed. We use the \texttt{NetworkX} library \cite{SciPyProceedings_11} to build them, ensuring that each sample represents a unique, non-isomorphic graph using the procedure described in Sec.~\ref{sec:graph-rep}.

For each graph family, we make sparse, medium, and dense instances with node counts $M \in \{8, 10, 14, 18\}$, corresponding to vectors of $28$, $45$, $91$, and $153$ bits, respectively. These categories refer to relative densities within a given node count and family; that is, for a fixed $M$ and graph type, sparse instances have lower average density than medium ones, and so on, though absolute values vary across families and sizes. Lastly, the size of each dataset depends on the number of nodes, as smaller graphs admit fewer unique configurations. Table \ref{tab:dataset-info} summarizes the properties of each dataset, including graph type, average density, percentage of bipartite graphs, average bipartivity, and number of samples. A detailed description of these metrics is provided in Sec.~\ref{sec:metrics}.

\subsection{Training}
The training objective is to minimize the MMD between the target and generated distributions. We train all models on a single laptop using the \texttt{IQPopt} library \cite{recio-armengolIQPoptFastOptimization2025}, an open-source tool for classically optimizing parameterized IQP circuits in JAX. We configure it with the ADAM optimizer \cite{kingmaAdamMethodStochastic2017}, the median heuristic for kernel bandwidth selection, and the data-driven initialization method of Recio-Armengol \textit{et al.}~\cite{recio-armengolTrainClassicalDeploy2025}, which sets two-qubit gate parameters proportional to the covariance of the corresponding bits in the training data. To refine these initial settings, we introduce scaling hyperparameters for both the kernel bandwidth and the initialization.

\subsection{Hyperparameter optimization}
To ensure a consistent basis for comparison across models, we perform HPO using the Optuna library \cite{optuna_2019} { employing a Tree-structured Parzen Estimator (TPE) sampler \cite{watanabeTreeStructuredParzenEstimator2025}. We allocate a budget of $60$ trials per dataset. This allocation is justifiable given that our search space has only three hyperparameters. Even under the conservative theoretical bounds of random search, $60$ iterations provide a $95\%$ probability ($1 - 0.95^{60} \approx 0.954$) of sampling a configuration within the top $5\%$ of the search space volume \cite{bergstraRandomSearchHyperParameter}. Since TPE leverages Bayesian optimization to prioritize promising regions, it is expected to locate optimal configurations more efficiently than this baseline.} Furthermore, each configuration is evaluated via repeated $k$-fold cross-validation, { with three repetitions and five folds. Once the best configuration is found, it is trained three separate times, with small perturbations on the initialization, and the best performing model is saved for sampling.}

We optimize three key parameters: the learning rate, which controls the optimizer step size; the bandwidth multiplier, which scales the median-heuristic kernel width; and the initialization multiplier, which adjusts the amplitude of the initial weights. Other settings---such as the optimizer choice---showed negligible influence on performance in preliminary tests and were therefore kept fixed.

\subsection{Sampling}
After training, we generate samples by preparing the $N$-qubit state with the optimized parameters on the IBM Aachen quantum computer and measuring in the computational basis. Each measurement produces an $N$-bit string $\mathbf{z}$, which we map to a graph following the procedure in Sec.~\ref{sec:graph-rep}. 

Importantly, we do not employ any error mitigation technique or classical post-processing, which allows us to directly assess the raw performance of the quantum hardware.

\subsection{Evaluation metrics}
\label{sec:metrics}

\begin{table*}[t]
  \centering
  \setlength{\tabcolsep}{3pt}
  \renewcommand{\arraystretch}{1.1}
    \caption{Performance validation for the $8$-node datasets ($28$-qubit models) conducted via simulations using PennyLane’s \texttt{lightning.qubit} simulator.}
  \label{tab:validation-performance}
  \begin{tabular}{c c *{3}{ c c }}
    \toprule
    Nodes & Type
      & \multicolumn{2}{c}{Dense}
      & \multicolumn{2}{c}{Medium}
      & \multicolumn{2}{c}{Sparse} \\
    \cmidrule(lr){3-4}
    \cmidrule(lr){5-6}
    \cmidrule(lr){7-8}
    & 
      & {$\mathbb E [\rho]$ (Difference)}
      & {BP.\% (Target \%)}
      & {$\mathbb E [\rho]$ (Difference)}
      & {BP.\% (Target \%)}
      & {$\mathbb E [\rho]$ (Difference)}
      & {BP.\% (Target \%)} \\
    \midrule
    \multirow{2}{*}{8}
      & BP
        & { 0.312 (-0.001) } & { 49.21 (100) }
        & { 0.230 (0.058) } & { 63.28 (100) }
        & { 0.228 (-0.003) } & { 77.19 (100) } \\
      & ER
        & { 0.759 (0.002) } & { 0.0 (0.0) }
        & { 0.437 (0.004) } & { 3.9 (0.5) }
        & { 0.221 (0.0005) } & { 56.64 (51.5) } \\    
    \bottomrule
  \end{tabular}
\end{table*}

We evaluate model performance by comparing the generated and target graph distributions across the features of interest.

For the density, we measure the deviation in the expected value between the generated and target distributions,
\begin{equation}
\Delta \mathbb E[\rho] =
\mathbb E_{\mathbf z\sim D_{\boldsymbol{\theta}}}[D(\mathbf z)] - \mathbb E_{\mathbf z\sim D}[D(\mathbf z)].
\end{equation} 
This test verifies whether the model reproduces the correct average edge density. Since this reflects the overall edge occupancy, a model that fails here will likely struggle to capture more complex features.

For the degree distribution, we employ a more comprehensive comparison. In an Erd\H{o}s–Rényi graph, where edges are independent, the degree distribution follows a binomial law. Accordingly, we compute the expected degree distribution from the generated samples and compare it to the corresponding binomial, both qualitatively---via histograms---and quantitatively using the total variation distance (TVD)
\begin{equation}
    \mathrm{TVD}(p, q_{\boldsymbol{\theta}}) =
    \frac{1}{2} \sum_k \big| p(k) - q_{\boldsymbol{\theta}}(k) \big|.
\end{equation}
Here $p(k)$ denotes the probability that a randomly selected node has degree $k$, and $q_{\boldsymbol{\theta}}(k)$ the corresponding probability under the model with parameters $\boldsymbol{\theta}$. This metric captures discrepancies in the full degree distribution, providing a sensitive test of how well local edge correlations are reproduced.

Lastly, we quantify the bipartite structure of the generated distributions using two complementary measures. First, bipartite accuracy, defined as the percentage of bipartite graphs generated, is determined via an exact $2$-coloring check. Given that this property is binary and can be very sensitive to noise---a single misplaced edge can introduce an odd-length cycle---we complement it with an expected bipartivity measure based on the spectral bipartivity index $\beta$ introduced by Estrada \textit{et al.} \cite{estradaSpectralMeasuresBipartivity2005}
\begin{equation}
    \beta =
    \frac{\sum_i \cosh{\lambda_i}}{\sum_i e^{\lambda_i}},
\end{equation}
where $\lambda_i$ are the eigenvalues of the adjacency matrix $\mathbf{A}$. This index compares the weighted contributions of even and odd length cycles: bipartite graphs yield $\beta=1$, while complete graphs yield $\beta=0$. 

We compute these metrics on $512$ samples per trained model. Simulations use PennyLane’s \texttt{lightning.qubit} backend~\cite{bergholmPennyLaneAutomaticDifferentiation2022}, while hardware demonstrations are executed on IBM’s $156$-qubit Aachen QPU, with each batch completing in $5.0 \pm 0.5$ seconds.

\section{Results}
\label{sec:results}

We organize the analysis in three stages. First, we validate the models at the $28$-qubit scale through noiseless simulations. Next, we assess their noise robustness by comparing the results of identical models from simulations and quantum hardware. Finally, we investigate scaling behavior by deploying models ranging from $28$ to $153$ qubits.

For each dataset, we {conduct a single HPO run and save the top $3$ best fitting models in terms of the cross-validation loss. Next, we sample to obtain the generation metrics.} From these, we select the best-performing model for each dataset according to the following criteria: for BP datasets, the model that generates the highest proportion of bipartite graphs; and for ER datasets, the model that minimizes the total variation distance of the degree distribution.

\subsection{Model validation}

{ In this section, we validate our approach by simulating the sampling of the $28$-qubit models in a noiseless regime. We evaluate performance based on the expected density---which corresponds to single-bodied correlations---and bipartite structure, which depends on all odd-order correlations. Together, these metrics probe the two ends of the expressivity spectrum. Successful modeling of the expected density implies the capacity to learn the lowest-order correlations. While the accuracy in capturing the bipartite structure examines whether the method is sufficiently expressive to represent complex, high-order dependencies.

All results are shown in Table~\ref{tab:validation-performance}. For ER datasets, we observe negligible deviations in expected density across all connectivity profiles, with a maximum error of $4 \times 10^{-3}$. In the context of $8$-node graphs ($28$ potential edges), this corresponds to an average overestimate of just $\sim 0.1$ edges, indicating that the mean edge probability is captured with high precision. To further confirm this finding, we calculate the TVD of the degree distributions, which provides a broader assessment of the accuracy. We find that the models maintain high precision across all densities, yielding TVD values of approximately $0.0075$ (Dense), $0.0138$ (Medium), and $0.0085$ (Sparse). Collectively, these results confirm that shallow IQP models are capable of reproducing first-order correlations with high precision in a noiseless environment.

For the BP datasets, we first distinguish learned structure from random occurrence. We employ the Erd\H{o}s–Rényi model as a null-hypothesis baseline: for each dataset, we generate $10^6$ random graphs matching the target density and measure the bipartite fraction. Performance exceeding this baseline indicates that the model can reproduce the bipartite structure beyond random chance.}

\begin{table}[t]
  \centering
  \setlength{\tabcolsep}{4pt}
  \renewcommand{\arraystretch}{1.1}
    \caption{Comparison of bipartite accuracies (\%) from BP models with those from an Erd\H{o}s–Rényi graph model of equal density. BP accuracies are computed from $512$ generated samples, while the baseline is obtained from $10^6$ random realizations.}
  \label{tab:bipartite_accuracies}
  \begin{tabular}{l c c c}
    \toprule
    \textbf{Density} & \textbf{Baseline} & \textbf{Generated} & \textbf{Improvement} \\
    \midrule
    Dense  & 25.159 & {49.218} & {+24.059} \\
    Medium & 32.163 & {63.281} & {+31.118} \\
    Sparse & 55.822 & {75.195} & {+19.373} \\
    \bottomrule
  \end{tabular}
\end{table}

{

As shown in Table~\ref{tab:bipartite_accuracies}, models across all densities significantly outperform the null hypothesis. The BP-Sparse model attains the best combined performance, reaching a bipartite accuracy of $75.19\%$---an improvement of nearly $20$ percentage points over the baseline---while maintaining an empirical density error of $-3\times 10^{-3}$ (Table~\ref{tab:validation-performance}). The BP-Medium and BP-Dense models demonstrate stronger performance on bipartite generation, achieving accuracies of $63.281\%$ and $49.218\%$ respectively, but fall shorter on the empirical density error.}


{ Collectively, these results show that in a noiseless $28$-qubit regime, shallow IQP circuits can effectively capture both local and global correlations. The models reproduce the expected density with high precision and bipartite structure beyond the null-hypothesis baseline. For broader context, we provide a comparison with classical generative models in Appendix \ref{app:classical_comparison}.}

\subsection{Model noise resilience} 

\begin{figure*}[htbp!]
    \centering
    \includegraphics[width=.8\linewidth]{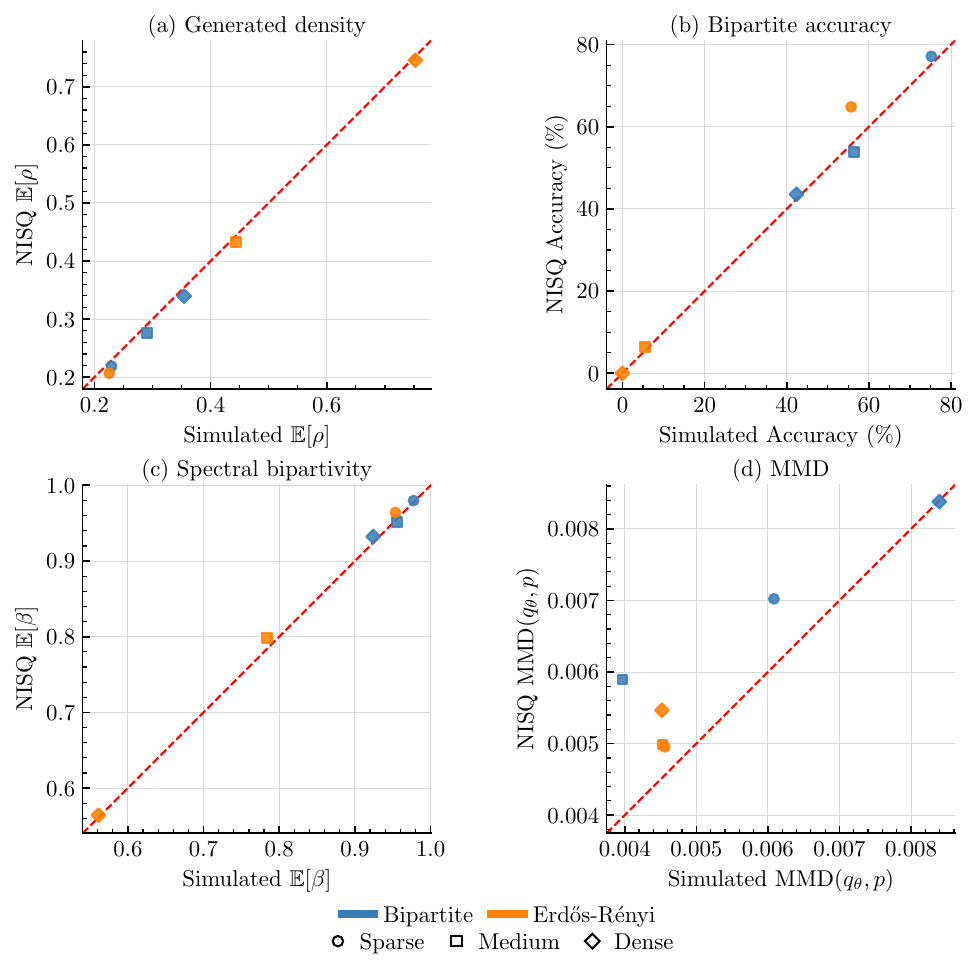}
    \caption{Comparison of graph generation metrics between NISQ hardware and classical simulations for identical models. 
    Panels show (a) generated density $\mathbb{E}[\rho]$, (b) bipartite accuracy, (c) spectral bipartivity $\mathbb{E}[\beta]$, (d) MMD$(q_\theta,p)$. The red dashed line denotes perfect agreement. Deviations highlight discrepancies due to hardware noise and finite sampling.
    }
    \label{fig:simvseval-all}
\end{figure*}

\begin{figure*}[htbp!]
    \centering
    \includegraphics[width=\linewidth]{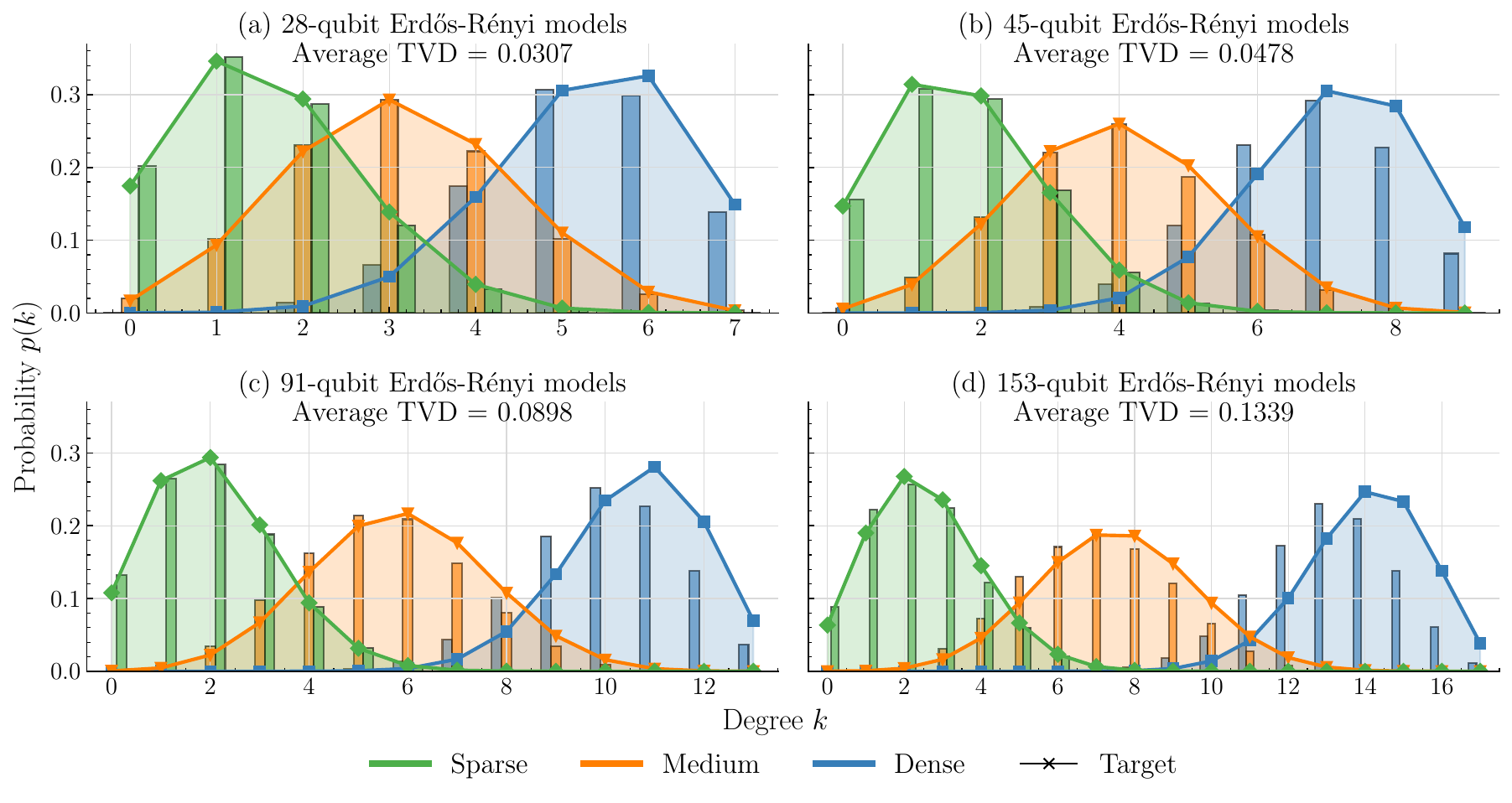}
    \caption{Degree distributions obtained from models from $28$ to $153$ qubits trained on Erd\H{o}s-Rényi datasets and executed on NISQ hardware. Bars indicate empirical node-degree frequencies, while solid lines denote theoretical binomial targets. The total variation distance (TVD) measures the deviation between generated and target distributions.}
    \label{fig:binomial-all}
\end{figure*}

Before scaling to regimes beyond classical simulation, we assess the noise resilience of our models by benchmarking $28$-qubit performance on IBM’s Aachen QPU against noiseless simulations. Consistency between backends would signal robustness to hardware noise, whereas significant deviations highlight sensitivity. { The quantum device characteristics and parameters at the time of data acquisition are outlined in Appendix \ref{app:device_characteristics}.}

{ Figure~\ref{fig:simvseval-all}(a) illustrates a near-perfect alignment between NISQ-measured densities and simulated values, confirming that single-bodied statistics are minimally affected by hardware noise at this scale. To further assess the impact of noise on such features, we compare the TVD of the degree distributions. While noiseless simulations yielded an average TVD of $\sim 0.01$, the quantum hardware achieves an average of $\sim 0.0307$ (see Fig.~\ref{fig:binomial-all}). This increase is expected due to hardware noise, yet the value remains low enough to confirm that local structures are robustly reproduced.

Remarkably, global features also exhibit notable resilience. Despite noise-induced errors, the bipartite accuracy on NISQ hardware (Fig.~\ref{fig:simvseval-all}(b)) closely tracks the noiseless ideal, with data points clustering tightly along the diagonal. This suggests that the learned bipartite structure is sufficiently robust for the noise profile of the quantum hardware at the $28$-qubit scale. Similarly, the expected spectral bipartivity (Fig.~\ref{fig:simvseval-all}(c)) remains in close agreement across backends. Finally, the MMD comparison (Fig.~\ref{fig:simvseval-all}(d)) corroborates this high precision; deviations between the results observed on simulations and quantum hardware are constrained to the order of $10^{-3}$.}




{Together, these observations suggest that while hardware noise impacts performance, it is not decisive at this scale. Local, single-body statistics exhibit the highest robustness, whereas global properties are comparatively more susceptible. However, contrary to the expectation that fragile global features would rapidly decay, we find that even strict bipartiteness is preserved with remarkable consistency at this scale.}

\subsection{Model scaling}

\begin{figure}[h!]
    \centering
    \includegraphics[width=\linewidth]{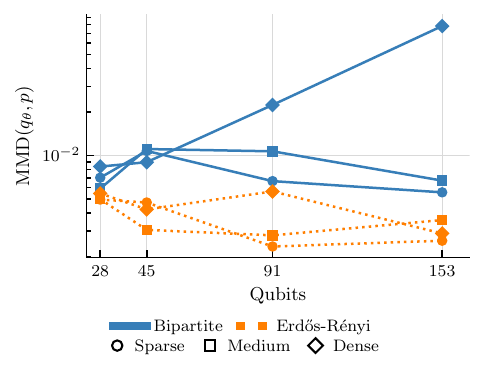}
    \caption{Scaling of the maximum mean discrepancy (MMD) across all models. MMD values are computed between samples produced on NISQ hardware and the target dataset.}
        \label{fig:scaling-mmd}
\end{figure}

\begin{figure*}[htbp!]
    \centering
    \includegraphics[width=.95\linewidth]{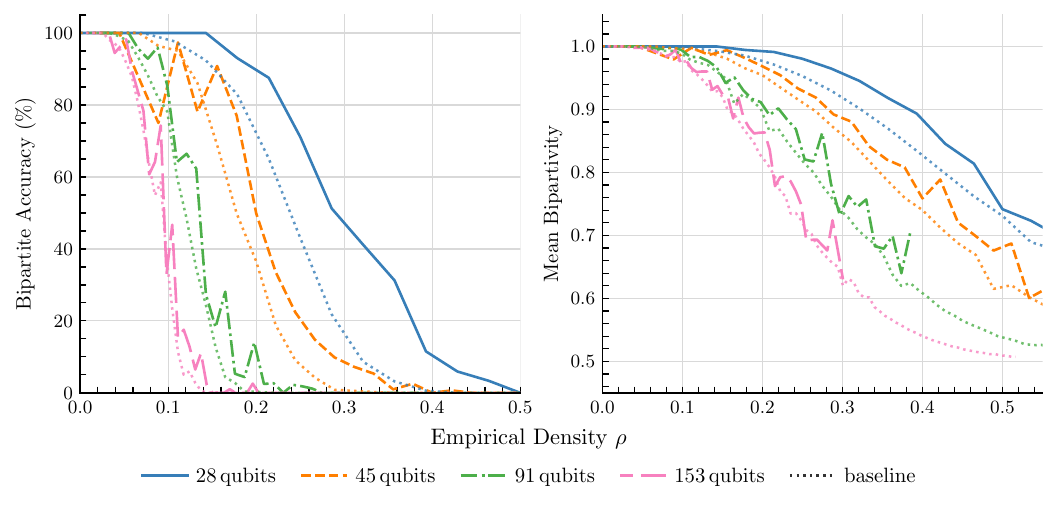}
    \caption{
    Scaling behavior of bipartite graph generation as a function of empirical density $\rho$ for models trained on bipartite datasets. Curves correspond to different qubit counts, with the dotted line showing the null-hypothesis baseline at each density.
    }
    \label{fig:scaling-bipartite}
\end{figure*}

Building on the validation and noise-resilience analysis, we now investigate the scaling behavior of shallow IQP models as they approach the limits of available quantum hardware. While the inherent imperfections of the NISQ regime constrain definitive conclusions regarding expressivity, these results establish an empirical baseline for generative performance on state-of-the-art devices.

{
We analyze the local connectivity of the ER models via the degree distribution and its TVD from the theoretical binomial (Fig.~\ref{fig:binomial-all}). The $28$- and $45$-qubit configurations reproduce the target statistics with high fidelity, yielding average TVDs of $0.0307$ and $0.0478$, respectively. However, deviation increases with system size: average TVDs rise to $0.0898$ at $91$ qubits and $0.1339$ at $153$ qubits, representing a $\sim3\times$ increase relative to the smallest instances. Given that the complexity of reproducing the degree distribution is independent of system size, we attribute the degradation primarily to noise. Despite this quantitative decline, qualitative performance remains consistent; as the histograms demonstrate, the generated distributions retain the characteristic binomial shape across all densities.

We employ the MMD as a global accuracy metric to track the divergence between learned and target distributions as system size increases (Fig.~\ref{fig:scaling-mmd}). The results reveal a clear hierarchy of learnability dictated by structural complexity. The Erd\H{o}s-Rényi families (orange) prove to be the easiest to learn, maintaining consistently low MMD values (below $10^{-2}$) across all scales. In contrast, the bipartite distributions (blue) exhibit higher divergence, particularly in the dense regime.

Notably, the MMD does not exhibit a monotonic increase with system size for BP models, as we would expect from the increased complexity at larger scales. For instance, the BP-Sparse configuration shows improving MMD values at larger scales, even as global features like strict bipartiteness degrade. This counterintuitive behavior underscores a critical limitation of using the MMD as a standalone metric in this context: the statistical correlations that dominate the loss landscape do not necessarily align with the defining structural features of the target distribution. Consequently, while a low MMD is a necessary condition for convergence, it is insufficient for validation; a rigorous assessment requires inspecting specific structural features of the generated data to fully characterize the model's fidelity.

For the BP models, we evaluate the learning of high-order correlations via the bipartite generation accuracy and spectral bipartivity (Fig.~\ref{fig:scaling-bipartite}). Both metrics exhibit a similar trend: performance degrades significantly as system size and density increase, reflecting the difficulty of maintaining global constraints at larger scales where noise is more pronounced. Strict bipartite accuracy drops to zero for the $91$- and $153$-qubit models shortly after the baseline, whereas smaller models consistently maintain higher performance. Notably, spectral bipartivity follows a similar decay but maintains a score exceeding the random baseline even when strict bipartite accuracy is zero (e.g., $\rho \approx 0.2-0.4$ at $91$ qubits). This indicates that partial bipartite features persist even if they are insufficient to satisfy the exact global constraint. Furthermore, the $91$- and $153$-qubit models fail to generate graphs beyond a specific connectivity threshold. This behavior suggests an optimization trade-off at large scales, unable to simultaneously reproduce the target density and enforce bipartiteness; the models appear to prioritize lower-density configurations where the constraint is less restrictive, effectively failing to populate the denser regions of the target distribution.}

Collectively, these results demonstrate that our approach maintains high fidelity for local features, such as degree distributions, even at the $153$-qubit scale. While global properties dependent on high-order correlations degrade with system size, the analysis of spectral bipartivity reveals that structural learning persists beyond the limits of strict bipartite accuracy. Ultimately, the models exhibit a clear hierarchy of learnability: unconstrained and local structures remain robust, whereas the combination of global features with strict constraints represents the current frontier for our approach on NISQ hardware.

\section{Conclusion}
\label{sec:conclusion}
{ 
In this work, we have introduced a method for quantum generative modeling on NISQ devices that combines the theoretical advantages of IQP circuits with a practical implementation strategy. By exploiting the classical trainability of the IQP ansatz, our approach bypasses the sampling overhead and noise constraints during training, enabling the optimization of large-scale models. Furthermore, we minimize the resources required for implementation while retaining classical hardness through an ansatz that maps to a 2D Ising model with a magnetic field.

We validated this approach through a formalized graph generation setup, which allowed us to probe the relationship between feature locality and model accuracy on simulations and real quantum hardware, ranging from $28$ to $153$ qubits. The proposed method demonstrated exceptional robustness in capturing local features, accurately reproducing the degree distribution of target graphs up to the full $153$-qubit scale without error mitigation. In contrast, global structural constraints such as bipartiteness were successfully learned and reproduced up to $45$ qubits—outperforming random baselines---but degraded at larger scales due to hardware noise.

These results position shallow IQP circuits as a viable, scalable solution for NISQ generative modeling. The ability to learn and generate distributions defined by local and intermediate-range correlations at over $100$ qubits suggests practical significance in domains where such features dominate, as our method is not limited to graph generation.}

\section*{DATA AVAILABILITY}
The code and data supporting the findings of this study are openly available in Zenodo \cite{ballo_gimbernat_2026_zenodo}.

\section*{Acknowledgments}
OB and MA are fellows of Eurecat’s ``Vicente L\'opez'' PhD grant program. This piece of research was carried out with the partial support of the following granted projects: SGR Grant 2021 SGR 01559 and RETECH EMT/43/2025 QML-CV from the Catalan Government, GRAIL PID2021-126808OB-I00 and from FEDER/UE, SUKIDI PID2024-157778OB-I00 grants from the Spanish Ministry of Science and Innovation, with the support of Cátedra UAB-Cruïlla grant TSI-100929-2023-2 from the Ministry of Economic Affairs and Digital Transformation of the Spanish Government. 

\bibliography{bib}

@book{newmanNetworks2018,
  title = {Networks},
  author = {Newman, Mark},
  date = {2018-10-18},
  year = {2018},
  volume = {1},
  publisher = {Oxford University Press},
  doi = {10.1093/oso/9780198805090.001.0001},
  url = {https://academic.oup.com/book/27884},
  urldate = {2025-06-04},
  isbn = {978-0-19-880509-0},
  langid = {english}
}

@article{drobyshevskiyRandomGraphModeling2020,
  title = {Random {{Graph Modeling}}: {{A Survey}} of the {{Concepts}}},
  shorttitle = {Random {{Graph Modeling}}},
  author = {Drobyshevskiy, Mikhail and Turdakov, Denis},
  year = {2020},
  month = nov,
  journal = {ACM Computing Surveys},
  volume = {52},
  number = {6},
  pages = {1--36},
  issn = {0360-0300, 1557-7341},
  doi = {10.1145/3369782},
  urldate = {2025-09-25},
  langid = {english}
}

@article{bonifatiGraphGeneratorsState2021,
  title = {Graph {{Generators}}: {{State}} of the {{Art}} and {{Open Challenges}}},
  shorttitle = {Graph {{Generators}}},
  author = {Bonifati, Angela and Holubov{\'a}, Irena and {Prat-P{\'e}rez}, Arnau and Sakr, Sherif},
  year = {2021},
  month = mar,
  journal = {ACM Computing Surveys},
  volume = {53},
  number = {2},
  pages = {1--30},
  issn = {0360-0300, 1557-7341},
  doi = {10.1145/3379445},
  urldate = {2025-09-25},
  langid = {english}
}

@article{bonginiMolecularGenerativeGraph2021,
  title = {Molecular Generative {{Graph Neural Networks}} for {{Drug Discovery}}},
  author = {Bongini, Pietro and Bianchini, Monica and Scarselli, Franco},
  year = {2021},
  month = aug,
  journal = {Neurocomputing},
  volume = {450},
  pages = {242--252},
  issn = {09252312},
  doi = {10.1016/j.neucom.2021.04.039},
  urldate = {2025-09-25},
  langid = {english}
}

@article{yangMoleculeGenerationDrug2024,
  title = {Molecule Generation for Drug Design: {{A}} Graph Learning Perspective},
  shorttitle = {Molecule Generation for Drug Design},
  author = {Yang, Nianzu and Wu, Huaijin and Zeng, Kaipeng and Li, Yang and Bao, Siyuan and Yan, Junchi},
  year = {2024},
  month = dec,
  journal = {Fundamental Research},
  pages = {S2667325824005259},
  issn = {26673258},
  doi = {10.1016/j.fmre.2024.11.027},
  urldate = {2025-09-25},
  langid = {english}
}

@inproceedings{cordeiroRandomGraphGeneration2010,
  title = {Random Graph Generation for Scheduling Simulations},
  booktitle = {Proceedings of the 3rd {{International ICST Conference}} on {{Simulation Tools}} and {{Techniques}}},
  author = {Cordeiro, Daniel and Mouni{\'e}, Gr{\'e}gory and Perarnau, Swann and Trystram, Denis and Vincent, Jean-Marc and Wagner, Fr{\'e}d{\'e}ric},
  year = {2010},
  publisher = {ICST},
  address = {Malaga, Spain},
  doi = {10.4108/ICST.SIMUTOOLS2010.8667},
  urldate = {2025-09-25},
  isbn = {978-963-9799-87-5},
  langid = {english}
}

@misc{guoSystematicSurveyDeep2022,
  title = {A {{Systematic Survey}} on {{Deep Generative Models}} for {{Graph Generation}}},
  author = {Guo, Xiaojie and Zhao, Liang},
  year = {2022},
  month = oct,
  number = {arXiv:2007.06686},
  eprint = {2007.06686},
  primaryclass = {cs},
  publisher = {arXiv},
  doi = {10.48550/arXiv.2007.06686},
  urldate = {2025-09-25},
  archiveprefix = {arXiv}
}

@misc{groverGraphiteIterativeGenerative2019,
  title = {Graphite: {{Iterative Generative Modeling}} of {{Graphs}}},
  shorttitle = {Graphite},
  author = {Grover, Aditya and Zweig, Aaron and Ermon, Stefano},
  year = {2019},
  month = may,
  number = {arXiv:1803.10459},
  eprint = {1803.10459},
  primaryclass = {stat},
  publisher = {arXiv},
  doi = {10.48550/arXiv.1803.10459},
  urldate = {2025-09-25},
  archiveprefix = {arXiv}
}

@misc{tranDeepNCDeepGenerative2020,
  title = {{{DeepNC}}: {{Deep Generative Network Completion}}},
  shorttitle = {{{DeepNC}}},
  author = {Tran, Cong and Shin, Won-Yong and Spitz, Andreas and Gertz, Michael},
  year = {2020},
  month = oct,
  number = {arXiv:1907.07381},
  eprint = {1907.07381},
  primaryclass = {cs},
  publisher = {arXiv},
  doi = {10.48550/arXiv.1907.07381},
  urldate = {2025-09-25},
  archiveprefix = {arXiv}
}

@article{bacciuEdgebasedSequentialGraph2020,
  title = {Edge-Based Sequential Graph Generation with Recurrent Neural Networks},
  author = {Bacciu, Davide and Micheli, Alessio and Podda, Marco},
  year = {2020},
  month = nov,
  journal = {Neurocomputing},
  volume = {416},
  pages = {177--189},
  issn = {09252312},
  doi = {10.1016/j.neucom.2019.11.112},
  urldate = {2025-09-25},
  langid = {english}
}

@misc{simonovskyGraphVAEGenerationSmall2018,
  title = {{{GraphVAE}}: {{Towards Generation}} of {{Small Graphs Using Variational Autoencoders}}},
  shorttitle = {{{GraphVAE}}},
  author = {Simonovsky, Martin and Komodakis, Nikos},
  year = {2018},
  month = feb,
  number = {arXiv:1802.03480},
  eprint = {1802.03480},
  primaryclass = {cs},
  publisher = {arXiv},
  doi = {10.48550/arXiv.1802.03480},
  urldate = {2025-09-25},
  archiveprefix = {arXiv}
}

@misc{flam-shepherdGraphDeconvolutionalGeneration2020,
  title = {Graph {{Deconvolutional Generation}}},
  author = {{Flam-Shepherd}, Daniel and Wu, Tony and {Aspuru-Guzik}, Alan},
  year = {2020},
  month = feb,
  number = {arXiv:2002.07087},
  eprint = {2002.07087},
  primaryclass = {cs},
  publisher = {arXiv},
  doi = {10.48550/arXiv.2002.07087},
  urldate = {2025-09-25},
  archiveprefix = {arXiv}
}

@misc{niuPermutationInvariantGraph2020,
  title = {Permutation {{Invariant Graph Generation}} via {{Score-Based Generative Modeling}}},
  author = {Niu, Chenhao and Song, Yang and Song, Jiaming and Zhao, Shengjia and Grover, Aditya and Ermon, Stefano},
  year = {2020},
  month = mar,
  number = {arXiv:2003.00638},
  eprint = {2003.00638},
  primaryclass = {cs},
  publisher = {arXiv},
  doi = {10.48550/arXiv.2003.00638},
  urldate = {2025-09-25},
  archiveprefix = {arXiv}
}

@misc{zoufalGenerativeQuantumMachine2021,
  title = {Generative {{Quantum Machine Learning}}},
  author = {Zoufal, Christa},
  year = {2021},
  month = nov,
  number = {arXiv:2111.12738},
  eprint = {2111.12738},
  primaryclass = {quant-ph},
  publisher = {arXiv},
  doi = {10.48550/arXiv.2111.12738},
  urldate = {2025-05-05},
  archiveprefix = {arXiv}
}

@article{shepherdInstantaneousQuantumComputation2009,
  title = {Instantaneous {{Quantum Computation}}},
  author = {Shepherd, Dan and Bremner, Michael J.},
  year = {2009},
  month = may,
  journal = {Proceedings of the Royal Society A: Mathematical, Physical and Engineering Sciences},
  volume = {465},
  number = {2105},
  eprint = {0809.0847},
  primaryclass = {quant-ph},
  pages = {1413--1439},
  issn = {1364-5021, 1471-2946},
  doi = {10.1098/rspa.2008.0443},
  urldate = {2025-05-12},
  archiveprefix = {arXiv}
}

@article{bremnerAverageCaseComplexityApproximate2016,
  title = {Average-{{Case Complexity Versus Approximate Simulation}} of {{Commuting Quantum Computations}}},
  author = {Bremner, Michael J. and Montanaro, Ashley and Shepherd, Dan J.},
  year = {2016},
  month = aug,
  journal = {Physical Review Letters},
  volume = {117},
  number = {8},
  pages = {080501},
  issn = {0031-9007, 1079-7114},
  doi = {10.1103/PhysRevLett.117.080501},
  urldate = {2025-05-16},
  copyright = {http://link.aps.org/licenses/aps-default-license},
  langid = {english}
}

@article{bremnerClassicalSimulationCommuting2011,
  title = {Classical Simulation of Commuting Quantum Computations Implies Collapse of the Polynomial Hierarchy},
  author = {Bremner, Michael J. and Jozsa, Richard and Shepherd, Dan J.},
  year = {2011},
  month = feb,
  journal = {Proceedings of the Royal Society A: Mathematical, Physical and Engineering Sciences},
  volume = {467},
  number = {2126},
  pages = {459--472},
  issn = {1364-5021, 1471-2946},
  doi = {10.1098/rspa.2010.0301},
  urldate = {2025-02-28},
  langid = {english}
}

@misc{marshallImprovedSeparationQuantum2024,
  title = {Improved Separation between Quantum and Classical Computers for Sampling and Functional Tasks},
  author = {Marshall, Simon C. and Aaronson, Scott and Dunjko, Vedran},
  year = {2024},
  month = oct,
  number = {arXiv:2410.20935},
  eprint = {2410.20935},
  primaryclass = {quant-ph},
  publisher = {arXiv},
  doi = {10.48550/arXiv.2410.20935},
  urldate = {2025-05-12},
  archiveprefix = {arXiv}
}

@article{nakataDiagonalQuantumCircuits2014,
  title = {Diagonal Quantum Circuits: Their Computational Power and Applications},
  shorttitle = {Diagonal Quantum Circuits},
  author = {Nakata, Yoshifumi and Murao, Mio},
  year = {2014},
  month = jul,
  journal = {The European Physical Journal Plus},
  volume = {129},
  number = {7},
  eprint = {1405.6552},
  primaryclass = {quant-ph},
  pages = {152},
  issn = {2190-5444},
  doi = {10.1140/epjp/i2014-14152-9},
  urldate = {2025-08-06},
  archiveprefix = {arXiv}
}

@misc{recio-armengolIQPoptFastOptimization2025,
  title = {{{IQPopt}}: {{Fast}} Optimization of Instantaneous Quantum Polynomial Circuits in {{JAX}}},
  shorttitle = {{{IQPopt}}},
  author = {{Recio-Armengol}, Erik and Bowles, Joseph},
  year = {2025},
  month = jan,
  number = {arXiv:2501.04776},
  eprint = {2501.04776},
  primaryclass = {quant-ph},
  publisher = {arXiv},
  doi = {10.48550/arXiv.2501.04776},
  urldate = {2025-05-05},
  archiveprefix = {arXiv}
}

@misc{nestSimulatingQuantumComputers2010,
  title = {Simulating Quantum Computers with Probabilistic Methods},
  author = {den Nest, M. Van},
  year = {2010},
  month = apr,
  number = {arXiv:0911.1624},
  eprint = {0911.1624},
  primaryclass = {quant-ph},
  publisher = {arXiv},
  doi = {10.48550/arXiv.0911.1624},
  urldate = {2025-05-05},
  archiveprefix = {arXiv}
}

@article{erdosRandomGraphs2022,
  title = {On Random Graphs. {{I}}.},
  author = {Erd{\H o}s, P. and R{\'e}nyi, A.},
  year = {2022},
  month = jul,
  journal = {Publicationes Mathematicae Debrecen},
  volume = {6},
  number = {3-4},
  pages = {290--297},
  issn = {00333883},
  doi = {10.5486/PMD.1959.6.3-4.12},
  urldate = {2025-05-13}
}

@article{JMLR:v13:gretton12a,
  author  = {Arthur Gretton and Karsten M. Borgwardt and Malte J. Rasch and Bernhard Sch{{\"o}}lkopf and Alexander Smola},
  title   = {A Kernel Two-Sample Test},
  journal = {Journal of Machine Learning Research},
  year    = {2012},
  volume  = {13},
  number  = {25},
  pages   = {723-773},
  url     = {http://jmlr.org/papers/v13/gretton12a.html}
}

@misc{sutherlandGenerativeModelsModel2021,
  title = {Generative {{Models}} and {{Model Criticism}} via {{Optimized Maximum Mean Discrepancy}}},
  author = {Sutherland, Danica J. and Tung, Hsiao-Yu and Strathmann, Heiko and De, Soumyajit and Ramdas, Aaditya and Smola, Alex and Gretton, Arthur},
  year = {2021},
  month = jan,
  number = {arXiv:1611.04488},
  eprint = {1611.04488},
  primaryclass = {stat},
  publisher = {arXiv},
  doi = {10.48550/arXiv.1611.04488},
  urldate = {2025-08-08},
  archiveprefix = {arXiv}
}

@article{rudolphTrainabilityBarriersOpportunities2024,
  title = {Trainability Barriers and Opportunities in Quantum Generative Modeling},
  author = {Rudolph, Manuel S. and Lerch, Sacha and Thanasilp, Supanut and Kiss, Oriel and Shaya, Oxana and Vallecorsa, Sofia and Grossi, Michele and Holmes, Zo{\"e}},
  year = {2024},
  month = nov,
  journal = {npj Quantum Information},
  volume = {10},
  number = {1},
  pages = {116},
  issn = {2056-6387},
  doi = {10.1038/s41534-024-00902-0},
  urldate = {2025-08-22},
  langid = {english}
}

@article{cerezoCostFunctionDependent2021,
  title = {Cost Function Dependent Barren Plateaus in Shallow Parametrized Quantum Circuits},
  author = {Cerezo, M. and Sone, Akira and Volkoff, Tyler and Cincio, Lukasz and Coles, Patrick J.},
  year = {2021},
  month = mar,
  journal = {Nature Communications},
  volume = {12},
  number = {1},
  pages = {1791},
  issn = {2041-1723},
  doi = {10.1038/s41467-021-21728-w},
  urldate = {2025-06-25},
  langid = {english}
}

@InProceedings{SciPyProceedings_11,
  author =       {Aric A. Hagberg and Daniel A. Schult and Pieter J. Swart},
  title =        {Exploring Network Structure, Dynamics, and Function using NetworkX},
  booktitle =   {Proceedings of the 7th Python in Science Conference},
  pages =     {11 - 15},
  address = {Pasadena, CA USA},
  year =      {2008},
  editor =    {Ga\"el Varoquaux and Travis Vaught and Jarrod Millman},
}

@misc{recio-armengolTrainClassicalDeploy2025,
  title = {Train on Classical, Deploy on Quantum: Scaling Generative Quantum Machine Learning to a Thousand Qubits},
  shorttitle = {Train on Classical, Deploy on Quantum},
  author = {{Recio-Armengol}, Erik and Ahmed, Shahnawaz and Bowles, Joseph},
  year = {2025},
  month = mar,
  number = {arXiv:2503.02934},
  eprint = {2503.02934},
  primaryclass = {quant-ph},
  publisher = {arXiv},
  doi = {10.48550/arXiv.2503.02934},
  urldate = {2025-03-07},
  archiveprefix = {arXiv},
  langid = {english}
}

@misc{kingmaAdamMethodStochastic2017,
  title = {Adam: {{A Method}} for {{Stochastic Optimization}}},
  shorttitle = {Adam},
  author = {Kingma, Diederik P. and Ba, Jimmy},
  year = {2017},
  month = jan,
  number = {arXiv:1412.6980},
  eprint = {1412.6980},
  primaryclass = {cs},
  publisher = {arXiv},
  doi = {10.48550/arXiv.1412.6980},
  urldate = {2025-06-04},
  archiveprefix = {arXiv}
}

@inproceedings{optuna_2019,
    title={Optuna: A Next-generation Hyperparameter Optimization Framework},
    author={Akiba, Takuya and Sano, Shotaro and Yanase, Toshihiko and Ohta, Takeru and Koyama, Masanori},
    booktitle={Proceedings of the 25th {ACM} {SIGKDD} International Conference on Knowledge Discovery and Data Mining},
    year={2019}
}

@article{fujiiQuantumCommutingCircuits2017,
  title = {Quantum {{Commuting Circuits}} and {{Complexity}} of {{Ising Partition Functions}}},
  author = {Fujii, Keisuke and Morimae, Tomoyuki},
  year = {2017},
  month = mar,
  journal = {New Journal of Physics},
  volume = {19},
  number = {3},
  eprint = {1311.2128},
  primaryclass = {quant-ph},
  pages = {033003},
  issn = {1367-2630},
  doi = {10.1088/1367-2630/aa5fdb},
  urldate = {2025-08-06},
  archiveprefix = {arXiv}
}

@article{mckayEfficientZGatesQuantum2017,
  title = {Efficient {{Z-Gates}} for {{Quantum Computing}}},
  author = {McKay, David C. and Wood, Christopher J. and Sheldon, Sarah and Chow, Jerry M. and Gambetta, Jay M.},
  year = {2017},
  month = aug,
  journal = {Physical Review A},
  volume = {96},
  number = {2},
  eprint = {1612.00858},
  primaryclass = {quant-ph},
  issn = {2469-9926, 2469-9934},
  doi = {10.1103/PhysRevA.96.022330},
  urldate = {2025-07-15},
  archiveprefix = {arXiv}
}

@misc{kurkinNoteUniversalityParameterized2025,
  title = {Note on the {{Universality}} of {{Parameterized IQP Circuits}} with {{Hidden Units}} for {{Generating Probability Distributions}}},
  author = {Kurkin, Andrii and Shen, Kevin and Pielawa, Susanne and Wang, Hao and Dunjko, Vedran},
  year = {2025},
  month = apr,
  number = {arXiv:2504.05997},
  eprint = {2504.05997},
  primaryclass = {quant-ph},
  publisher = {arXiv},
  doi = {10.48550/arXiv.2504.05997},
  urldate = {2025-05-05},
  archiveprefix = {arXiv}
}

@article{laroccaBarrenPlateausVariational2025,
  title = {Barren Plateaus in Variational Quantum Computing},
  author = {Larocca, Mart{\'i}n and Thanasilp, Supanut and Wang, Samson and Sharma, Kunal and Biamonte, Jacob and Coles, Patrick J. and Cincio, Lukasz and McClean, Jarrod R. and Holmes, Zo{\"e} and Cerezo, M.},
  year = {2025},
  month = mar,
  journal = {Nature Reviews Physics},
  volume = {7},
  number = {4},
  pages = {174--189},
  issn = {2522-5820},
  doi = {10.1038/s42254-025-00813-9},
  urldate = {2025-06-25},
  langid = {english}
}

@article{cerezoDoesProvableAbsence2025,
  title = {Does Provable Absence of Barren Plateaus Imply Classical Simulability?},
  author = {Cerezo, M. and Larocca, Martin and {Garc{\'i}a-Mart{\'i}n}, Diego and Diaz, N. L. and Braccia, Paolo and Fontana, Enrico and Rudolph, Manuel S. and Bermejo, Pablo and Ijaz, Aroosa and Thanasilp, Supanut and Anschuetz, Eric R. and Holmes, Zo{\"e}},
  year = {2025},
  month = aug,
  journal = {Nature Communications},
  volume = {16},
  number = {1},
  pages = {7907},
  issn = {2041-1723},
  doi = {10.1038/s41467-025-63099-6},
  urldate = {2025-08-27},
  langid = {english}
}

@misc{ballo_gimbernat_2026_zenodo,
  author       = {Balló-Gimbernat, O. and Arroyo-Sánchez, M. and García Molina, P. and Garriga, A. and Vilariño, F.},
  title        = {Code for "{Shallow} instantaneous quantum polynomial-time circuits for generative modeling on noisy intermediate-scale quantum hardware"},
  year         = {2026},
  publisher    = {Zenodo},
  doi          = {10.5281/zenodo.18983816},
  note         = {Data available at \url{https://doi.org/10.5281/zenodo.18983816}}
}

@article{bergstraRandomSearchHyperParameter,
  title = {Random {{Search}} for {{Hyper-Parameter Optimization}}},
  author = {Bergstra, James and Bergstra, James and Bengio, Yoshua and Bengio, Yoshua},
  journal = {Journal of Machine Learning Research},
  langid = {english}
}

@misc{watanabeTreeStructuredParzenEstimator2025,
  title = {Tree-{{Structured Parzen Estimator}}: {{Understanding Its Algorithm Components}} and {{Their Roles}} for {{Better Empirical Performance}}},
  shorttitle = {Tree-{{Structured Parzen Estimator}}},
  author = {Watanabe, Shuhei},
  year = 2025,
  month = sep,
  number = {arXiv:2304.11127},
  eprint = {2304.11127},
  primaryclass = {cs},
  publisher = {arXiv},
  doi = {10.48550/arXiv.2304.11127},
  urldate = {2026-02-12},
  archiveprefix = {arXiv}
}

@article{kastureProtocolsClassicallyTraining2023,
  title = {Protocols for Classically Training Quantum Generative Models on Probability Distributions},
  author = {Kasture, Sachin and Kyriienko, Oleksandr and Elfving, Vincent E.},
  year = 2023,
  month = oct,
  journal = {Physical Review A},
  volume = {108},
  number = {4},
  eprint = {2210.13442},
  primaryclass = {quant-ph},
  pages = {042406},
  issn = {2469-9926, 2469-9934},
  doi = {10.1103/PhysRevA.108.042406},
  urldate = {2025-11-19},
  archiveprefix = {arXiv}
}

@article{coyleBornSupremacyQuantum2020,
  title = {The {{Born}} Supremacy: Quantum Advantage and Training of an {{Ising Born}} Machine},
  shorttitle = {The {{Born}} Supremacy},
  author = {Coyle, Brian and Mills, Daniel and Danos, Vincent and Kashefi, Elham},
  year = {2020},
  month = jul,
  journal = {npj Quantum Information},
  volume = {6},
  number = {1},
  pages = {60},
  issn = {2056-6387},
  doi = {10.1038/s41534-020-00288-9},
  urldate = {2025-08-05},
  langid = {english}
}

@article{liuDifferentiableLearningQuantum2018,
  title = {Differentiable {{Learning}} of {{Quantum Circuit Born Machine}}},
  author = {Liu, Jin-Guo and Wang, Lei},
  year = {2018},
  month = dec,
  journal = {Physical Review A},
  volume = {98},
  number = {6},
  eprint = {1804.04168},
  primaryclass = {quant-ph},
  pages = {062324},
  issn = {2469-9926, 2469-9934},
  doi = {10.1103/PhysRevA.98.062324},
  urldate = {2025-04-18},
  archiveprefix = {arXiv}
}

@misc{bergholmPennyLaneAutomaticDifferentiation2022,
  title = {{{PennyLane}}: {{Automatic}} Differentiation of Hybrid Quantum-Classical Computations},
  shorttitle = {{{PennyLane}}},
  author = {Bergholm, Ville and Izaac, Josh and Schuld, Maria and Gogolin, Christian and Ahmed, Shahnawaz and Ajith, Vishnu and Alam, M. Sohaib and {Alonso-Linaje}, Guillermo and AkashNarayanan, B. and Asadi, Ali and Arrazola, Juan Miguel and Azad, Utkarsh and Banning, Sam and Blank, Carsten and Bromley, Thomas R. and Cordier, Benjamin A. and Ceroni, Jack and Delgado, Alain and Matteo, Olivia Di and Dusko, Amintor and Garg, Tanya and Guala, Diego and Hayes, Anthony and Hill, Ryan and Ijaz, Aroosa and Isacsson, Theodor and Ittah, David and Jahangiri, Soran and Jain, Prateek and Jiang, Edward and Khandelwal, Ankit and Kottmann, Korbinian and Lang, Robert A. and Lee, Christina and Loke, Thomas and Lowe, Angus and McKiernan, Keri and Meyer, Johannes Jakob and {Monta{\~n}ez-Barrera}, J. A. and Moyard, Romain and Niu, Zeyue and O'Riordan, Lee James and Oud, Steven and Panigrahi, Ashish and Park, Chae-Yeun and Polatajko, Daniel and Quesada, Nicol{\'a}s and Roberts, Chase and S{\'a}, Nahum and Schoch, Isidor and Shi, Borun and Shu, Shuli and Sim, Sukin and Singh, Arshpreet and Strandberg, Ingrid and Soni, Jay and Sz{\'a}va, Antal and Thabet, Slimane and {Vargas-Hern{\'a}ndez}, Rodrigo A. and Vincent, Trevor and Vitucci, Nicola and Weber, Maurice and Wierichs, David and Wiersema, Roeland and Willmann, Moritz and Wong, Vincent and Zhang, Shaoming and Killoran, Nathan},
  year = {2022},
  month = jul,
  number = {arXiv:1811.04968},
  eprint = {1811.04968},
  primaryclass = {quant-ph},
  publisher = {arXiv},
  doi = {10.48550/arXiv.1811.04968},
  urldate = {2025-05-08},
  archiveprefix = {arXiv}
}

@article{swekeQuantumClassicalLearnability2021,
  title = {On the {{Quantum}} versus {{Classical Learnability}} of {{Discrete Distributions}}},
  author = {Sweke, Ryan and Seifert, Jean-Pierre and Hangleiter, Dominik and Eisert, Jens},
  year = {2021},
  month = mar,
  journal = {Quantum},
  volume = {5},
  eprint = {2007.14451},
  primaryclass = {quant-ph},
  pages = {417},
  issn = {2521-327X},
  doi = {10.22331/q-2021-03-23-417},
  urldate = {2025-06-30},
  archiveprefix = {arXiv}
}

@article{estradaSpectralMeasuresBipartivity2005,
  title = {Spectral Measures of Bipartivity in Complex Networks},
  author = {Estrada, Ernesto and {Rodr{\'i}guez-Vel{\'a}zquez}, Juan A.},
  year = {2005},
  month = oct,
  journal = {Physical Review E},
  volume = {72},
  number = {4},
  pages = {046105},
  issn = {1539-3755, 1550-2376},
  doi = {10.1103/PhysRevE.72.046105},
  urldate = {2025-10-15},
  copyright = {http://link.aps.org/licenses/aps-default-license},
  langid = {english}
}

@article{kingmaIntroductionVariationalAutoencoders2019,
  title = {An {{Introduction}} to {{Variational Autoencoders}}},
  author = {Kingma, Diederik P. and Welling, Max},
  year = 2019,
  month = nov,
  journal = {Foundations and Trends\textregistered{} in Machine Learning},
  volume = {12},
  number = {4},
  eprint = {1906.02691},
  primaryclass = {cs},
  pages = {307--392},
  issn = {1935-8237, 1935-8245},
  doi = {10.1561/2200000056},
  urldate = {2026-01-13},
  archiveprefix = {arXiv}
}

@article{kissConditionalBornMachine2022,
  title = {Conditional {{Born}} Machine for {{Monte Carlo}} Event Generation},
  author = {Kiss, Oriel and Grossi, Michele and Kajomovitz, Enrique and Vallecorsa, Sofia},
  year = 2022,
  month = aug,
  journal = {Physical Review A},
  volume = {106},
  number = {2},
  pages = {022612},
  issn = {2469-9926, 2469-9934},
  doi = {10.1103/PhysRevA.106.022612},
  urldate = {2026-01-14},
  langid = {english}
}

@article{liuDifferentiableLearningQuantum2018a,
  title = {Differentiable Learning of Quantum Circuit {{Born}} Machines},
  author = {Liu, Jin-Guo and Wang, Lei},
  year = 2018,
  month = dec,
  journal = {Physical Review A},
  volume = {98},
  number = {6},
  pages = {062324},
  issn = {2469-9926, 2469-9934},
  doi = {10.1103/PhysRevA.98.062324},
  urldate = {2026-01-21},
  langid = {english}
}

@article{mitaraiQuantumCircuitLearning2018a,
  title = {Quantum Circuit Learning},
  author = {Mitarai, K. and Negoro, M. and Kitagawa, M. and Fujii, K.},
  year = 2018,
  month = sep,
  journal = {Physical Review A},
  volume = {98},
  number = {3},
  pages = {032309},
  issn = {2469-9926, 2469-9934},
  doi = {10.1103/PhysRevA.98.032309},
  urldate = {2026-01-21},
  langid = {english}
}

@article{zengLearningInferenceGenerative2019,
  title = {Learning and Inference on Generative Adversarial Quantum Circuits},
  author = {Zeng, Jinfeng and Wu, Yufeng and Liu, Jin-Guo and Wang, Lei and Hu, Jiangping},
  year = 2019,
  month = may,
  journal = {Physical Review A},
  volume = {99},
  number = {5},
  pages = {052306},
  issn = {2469-9926, 2469-9934},
  doi = {10.1103/PhysRevA.99.052306},
  urldate = {2026-01-21},
  langid = {english}
}

\onecolumngrid
\appendix
\section{Derivation of the maximum mean discrepancy as a quantum observable}
\label{appendix_mmd}
{
This appendix details the derivation of the squared Maximum Mean Discrepancy (MMD) as a mixture of Pauli-Z expectation values. This formulation, first introduced in Rudolph \textit{et al.} 2024 \cite{rudolphTrainabilityBarriersOpportunities2024}, allows the classical training of Instantaneous Quantum Polynomial (IQP) circuits as implicit generative models.

The squared MMD is an integral probability metric that quantifies the distance between a target distribution $p$ and a model distribution $q_\theta$. It is defined via a kernel $k(\boldsymbol x, \boldsymbol y)$ as,

\begin{equation}
\begin{split}
    \text{MMD}^2(p,q_\theta) &= \mathbb E_{\boldsymbol{x}\sim p,\boldsymbol y\sim p}[k(\boldsymbol x,\boldsymbol y)] \\
               &\quad - 2\mathbb E_{\boldsymbol x\sim p, \boldsymbol y\sim q_\theta}[k(\boldsymbol x,\boldsymbol y)] \\
               &\quad + \mathbb E_{\boldsymbol x\sim q_\theta, \boldsymbol y\sim q_\theta}[k(\boldsymbol x,\boldsymbol y)].
\end{split}
\end{equation}

Crucial to this derivation, we employ the Gaussian kernel,

\begin{equation}
    k_\sigma(\boldsymbol x, \boldsymbol y) = \exp{-\frac{||\boldsymbol x-\boldsymbol y||^2}{2\sigma^2}}.
\end{equation}

Our goal is to rewrite the Euclidean distance $||\boldsymbol x - \boldsymbol y||^2$ with quantum operators. First, since $\boldsymbol x$ and $\boldsymbol y$ are binary strings, the expression becomes the Hamming distance,

\begin{equation}
||\boldsymbol x - \boldsymbol y ||^2=\sum_{i=1}^n(x_i - y_i)^2.
\end{equation}

Next, through the properties of the exponential, the kernel becomes

\begin{equation}
k(\boldsymbol x, \boldsymbol y)=\prod_{i=1}^n\exp-\frac{(x_i-y_i)^2}{2\sigma^2},
\end{equation}

since $e^{a+b}=e^{a}e^b$. Then we can write the application of the kernel on a single bit as,

\begin{equation}
k_i( x_i, y_i)=\exp-\frac{(x_i-y_i)^2}{2\sigma^2},
\end{equation}

which is $1$ whenever $x_i=y_i$ and $e^{-1/2\sigma^2}$ otherwise. We seek an operator that satisfies those same conditions. In Rudolph \textit{et al.} 2024 \cite{rudolphTrainabilityBarriersOpportunities2024}, they show that

\begin{equation}
k_i( x_i, y_i)=\exp-\frac{(x_i-y_i)^2}{2\sigma^2}\equiv(1-p_\sigma) + p_\sigma(-1)^{x_i}(-1)^{y_i},
\end{equation}

where $p_\sigma$ is a scalar parameter that we can obtain by enforcing the previous conditions. For $x_i=y_i$ we trivially obtain $(1-p_\sigma) +p_\sigma=1$. For $x_i\ne y_i$ we obtain $(1-p_\sigma) -p_\sigma=e^{-1/2\sigma^2}$, which yields

\begin{equation}
p_\sigma=\frac{1-e^{-\frac{1}{2\sigma^2}}}{2}.
\end{equation}

We observe that the term $(-1)^{x_i}(-1)^{y_i}$ corresponds to the eigenvalue of the Pauli-Z tensor product $Z_i \otimes Z_i$ acting on the computational basis states $|x_i\rangle \otimes |y_i\rangle$. Consequently, the single-bit kernel is equivalent to the expectation value of the quantum operator

\begin{equation}
O_i= (1-p_\sigma)\mathbb I\otimes \mathbb I + p_\sigma(Z_i \otimes Z_i).
\end{equation}

The full kernel operator is then the tensor product of these single-qubit operators across all n qubits,

\begin{equation}
\hat K=\bigotimes_{i=1}^{n}\left[ (1-p_\sigma)\mathbb I\otimes \mathbb I + p_\sigma(Z_i \otimes Z_i)\right].
\end{equation}

Expanding this tensor product yields a sum over all possible Pauli-Z strings. We define $Z_{\boldsymbol a}=\bigotimes_{i=1}^{n}Z_i^{a_i}$ for a bitstring $\boldsymbol a\in \{0,1\}^n$, where the operator $Z_i$ is applied if and only if $a_i=1$. The weight of each term in the expansion is determined by the number of $Z$ operators chosen (the Hamming weight $|\boldsymbol a|$). This results in the operator expansion:

\begin{equation}
\hat K =\sum_{\boldsymbol a \in\{0,1\}^n}\mathcal P_\sigma(\boldsymbol a)(Z_{\boldsymbol a}\otimes Z_{\boldsymbol a}),
\end{equation}

where $\mathcal P_\sigma(\boldsymbol a)$ corresponds to the probability of sampling the string $\boldsymbol a$ from a product of Bernoulli distributions,

\begin{equation}
\mathcal P_\sigma(\boldsymbol a)=(1-p_\sigma)^{n-|\boldsymbol a|}p_\sigma^{|\boldsymbol a|}.
\end{equation}

Returning to the classical domain, the kernel evaluation is given by the expectation of this operator with respect to the basis states $|\boldsymbol x\rangle$ and $|\boldsymbol y\rangle$, $\langle \boldsymbol x, \boldsymbol y | \hat K | \boldsymbol x, \boldsymbol y \rangle$. We note that the eigenvalue of the operator $Z_{\boldsymbol a}$ on a computational basis state is the product of the individual bitwise eigenvalues,

\begin{equation}
Z_{\boldsymbol a}|\boldsymbol x\rangle = (-1)^{\boldsymbol x\cdot \boldsymbol a}|\boldsymbol x\rangle
\end{equation}

Substituting this eigenvalue into the operator expansion, the kernel function becomes,

\begin{equation}
k(\boldsymbol x, \boldsymbol y) = \sum_{\boldsymbol a \in \{0,1\}^n} \mathcal{P}_\sigma(\boldsymbol a) (-1)^{\boldsymbol x \cdot \boldsymbol a} (-1)^{\boldsymbol y \cdot \boldsymbol a}.
\end{equation}

Finally, we substitute this expansion into the definition of the MMD. By linearity of expectation, we can interchange the sum and the expectations. For the cross-term, this yields

\begin{equation}
\mathbb E_{\boldsymbol x\sim p, \boldsymbol y\sim q_\theta}[k(\boldsymbol x,\boldsymbol y)] = \sum_{\boldsymbol a} \mathcal{P}_\sigma(\boldsymbol a) \mathbb E_{\boldsymbol x\sim p}[(-1)^{\boldsymbol x \cdot \boldsymbol a}] \mathbb E_{\boldsymbol y\sim q_\theta}[(-1)^{\boldsymbol y \cdot \boldsymbol a}].
\end{equation}

Identifying the inner expectations as the moments $\langle Z_{\boldsymbol a}\rangle_p$ and $\langle Z_{\boldsymbol a}\rangle_{q_\theta}$ respectively, and applying the same logic to the self-similarity terms, the MMD simplifies to a weighted sum of squared differences between expectations,

\begin{equation}
\text{MMD}^2(p,q_\theta) = \mathbb E_{\boldsymbol a \sim \mathcal{P}_\sigma} \left[ \left( \langle Z_{\boldsymbol a} \rangle_p - \langle Z_{\boldsymbol a} \rangle_{q_\theta} \right)^2 \right].
\end{equation}
}
\section{Comparison with classical models}
\label{app:classical_comparison}

{The classical training of IQP circuits naturally raises questions regarding whether this hybrid approach compromises potential quantum advantages or if it simply reproduces results achievable by standard classical generative models. To address this, we compare the performance of the proposed IQP models against a GraphVAE~\cite{simonovskyGraphVAEGenerationSmall2018} baseline for both bipartite and Erd\H{o}s–Rényi (ER) graph generation.

\subsection{GraphVAE Architecture}
The GraphVAE~\cite{simonovskyGraphVAEGenerationSmall2018} is a deep generative model specifically designed to generate small graphs (between $9$ and $38$ nodes). It formulates the generation process within a Variational Autoencoder (VAE)~\cite{kingmaIntroductionVariationalAutoencoders2019} that outputs a probabilistic, fully-connected graph of a predefined maximum size in a single shot, similar to the sampling of quantum circuits.

The model consists of a graph encoder $q_\phi(\mathbf{z}|G)$ and a deterministic decoder $p_\theta(G|\mathbf{z})$. While the original architecture supports graphs with variable node counts and features $H$, we simplify the model for our fixed-size, unweighted, undirected datasets. We omit node feature vectors and, since the node count is fixed to $N$, we remove the node-existence predictions (originally encoded on the diagonal) by enforcing a zero diagonal ($\tilde{A}_{ii} = 0$) in the output adjacency matrix $A \in \{0,1\}^{N \times N}$.

The encoder utilizes a feed-forward network employing Edge-Conditioned Graph Convolutions (ECC)~\cite{simonovskyGraphVAEGenerationSmall2018} to embed the input graph into a continuous latent representation. This process propagates node information through the graph layers, which is then aggregated into a fixed-size graph representation using a gated global pooling mechanism. This aggregated representation parameterizes the variational posterior $q_\phi(\mathbf{z}|G)$ (assumed to be a multivariate Gaussian), from which the latent vector $\mathbf{z} \in \mathbb{R}^{d_z}$ is sampled via the reparameterization trick.

The decoder is structured as a multi-layer perceptron (MLP) that maps the latent vector $\mathbf{z}$ directly to a probabilistic adjacency matrix $\tilde{A} \in [0,1]^{N \times N}$. The entries $\tilde{A}_{i,j}$ represent the probability of an edge existing between nodes $i$ and $j$. The model assumes that these edge probabilities are Bernoulli variables that are independent conditioned on $\mathbf{z}$. Crucially, this independence assumption does not limit the model's ability to capture global structure (such as bipartiteness). The complex correlations required for such global features are encoded within the latent variable $\mathbf{z}$ itself; the decoder simply acts as a mapping from this globally correlated latent state to the individual edge probabilities.

\subsection{Training Objective and Graph Matching}
A fundamental challenge in graph generation tackled by this model is the lack of a canonical node ordering; a generated adjacency matrix $\hat{A}$ may be isomorphic to the ground truth $A$ under a node permutation $P$, yet appear distinct in a standard element-wise comparison. To evaluate the reconstruction likelihood $p_{\theta}(G|\mathbf{z})$, the model must align the generated probabilistic graph with the ground truth.

The GraphVAE optimizes an Evidence Lower Bound (ELBO) objective, comprising a reconstruction loss and a Kullback-Leibler (KL) divergence term. While the original architecture employs approximate graph matching (specifically Max-Pooling Matching) to scale to larger graphs, the small system size considered in this comparison ($N=8$) allows for tractable exact matching. We define the reconstruction loss by explicitly iterating over the set of all possible permutation matrices $\mathcal{P}_N$ to find the optimal alignment,
\begin{equation}
    \mathcal{L}_{\text{recon}} = \min_{P \in \mathcal{P}_N} \mathcal{L}_{\text{BCE}}(\hat{A}, P A P^T),
\end{equation}
where $\mathcal{L}_{\text{BCE}}$ denotes the binary cross-entropy loss. This ensures that the training objective is invariant to the arbitrary node ordering of the dataset.

\subsection{Results}
The GraphVAE models were subjected to hyperparameter optimization (HPO) with a computational budget comparable to that of the shallow IQP circuits. We define this budget as the total number of optimization steps across all HPO trials (approximately $10^4$ total epochs for both model classes). The GraphVAE search space swept over the latent space dimension $d_z \in [16, 64]$, learning rate $\eta \in [10^{-4}, 10^{-2}]$, and the KL annealing rate $\beta\in[10^{-3}, 1.0]$. Table~\ref{tab:bip-vae-metrics} presents the performance comparison for bipartite generation accuracy. While the GraphVAE models generally achieve higher accuracy in capturing the global bipartite structure---benefiting from their high parameter count and deep inductive biases---shallow IQP circuits remain competitive, particularly given the orders-of-magnitude difference in model complexity.

\begin{table}[h!]
  \centering
  \renewcommand{\arraystretch}{1.5}
  \setlength{\tabcolsep}{10pt}
  \caption{Comparison of bipartite generation accuracy between shallow IQP circuits and GraphVAE models trained on $8$-node bipartite datasets.}
  \label{tab:bip-vae-metrics}
  \begin{tabular}{llc}
    \toprule
    \textbf{Density} & \textbf{Model} & \textbf{Bip (\%)} \\
    \midrule
    \multirow{2}{*}{Sparse} & GraphVAE & \textbf{86.5} \\
                            & IQP      & 75.19 \\
    \midrule
    \multirow{2}{*}{Medium} & GraphVAE & \textbf{86.1} \\
                            & IQP      & 63.3 \\
    \midrule
    \multirow{2}{*}{Dense}  & GraphVAE & \textbf{80.9} \\
                            & IQP      & 49.2 \\
    \bottomrule
  \end{tabular}
\end{table}

Interestingly, for Erd\H{o}s–Rényi (ER) datasets, the performance gap narrows significantly. As shown in Table~\ref{tab:er-vae-metrics}, shallow IQP circuits outperform the GraphVAE baseline in reproducing the target average edge probability $\rho$ in all three cases.

\begin{table}[h!]
  \centering
  \renewcommand{\arraystretch}{1.5}
  \setlength{\tabcolsep}{9pt}
  \caption{Comparison of generated average edge probability between shallow IQP circuits and GraphVAE models trained on $8$-node ER datasets.}
  \label{tab:er-vae-metrics}
  \begin{tabular}{llccc}
    \toprule
    \textbf{Density} & \textbf{Model} & \textbf{$\rho$} & \textbf{Target $\rho$} \\
    \midrule
    \multirow{2}{*}{Sparse} & GraphVAE  & 0.2242 & \multirow{2}{*}{0.2207} \\
                            & IQP       & \textbf{0.2204} & \\
    \midrule
    \multirow{2}{*}{Medium} & GraphVAE & 0.4308 & \multirow{2}{*}{0.4420} \\
                            & IQP      & \textbf{0.4419} & \\
    \midrule
    \multirow{2}{*}{Dense}  & GraphVAE  & 0.7543 & \multirow{2}{*}{0.7618} \\
                            & IQP       & \textbf{0.7617} & \\
    \bottomrule
  \end{tabular}
\end{table}

To contextualize these results, Table~\ref{tab:param_comparison} compares the parameter counts across models against the performance in terms of the Total Variation Distance (TVD) of the generated degree distributions. The shallow IQP circuits achieve comparable or superior performance on these local features while utilizing approximately $3,500\times$ fewer parameters ($56$ vs. $\approx 200,000$).

\begin{table}[h!]
\centering
\renewcommand{\arraystretch}{1.5}
\setlength{\tabcolsep}{7pt}
\caption{Comparison of parameter count and TVD of the generated degree distributions across models trained on $8$-node datasets.}
\label{tab:param_comparison}
\begin{tabular}{lllcc}
\toprule
\textbf{Graph} & \textbf{Density} & \textbf{Model} & \textbf{Params} & \textbf{TVD} \\
\midrule
\multirow{6}{*}{Bipartite} & \multirow{2}{*}{Sparse} & GraphVAE & 205.2k & 0.0811 \\
 & & IQP & \textbf{56} & \textbf{0.0189} \\
 \cmidrule(l){2-5}

 & \multirow{2}{*}{Medium} & GraphVAE & 200.1k & \textbf{0.0629} \\
 & & IQP & \textbf{56} & 0.1428 \\
 \cmidrule(l){2-5}

 & \multirow{2}{*}{Dense} & GraphVAE & 199.4k & 0.0789 \\
 & & IQP & \textbf{56} & \textbf{0.0357} \\
\midrule
\multirow{6}{*}{ER} & \multirow{2}{*}{Sparse} & GraphVAE & 200.1k & {0.0242} \\
 & & IQP & \textbf{56} & \textbf{0.0085} \\
 \cmidrule(l){2-5}

 & \multirow{2}{*}{Medium} & GraphVAE & 202.8k & 0.0597 \\
 & & IQP & \textbf{56} & \textbf{0.0137} \\
 \cmidrule(l){2-5}

 & \multirow{2}{*}{Dense} & GraphVAE & 206.7k & 0.0382 \\
 & & IQP & \textbf{56} & \textbf{0.0074} \\
\bottomrule
\end{tabular}
\end{table}

We emphasize that this comparison is not intended to demonstrate ``quantum advantage'' in the strict computational sense. Rather, it serves as an empirical benchmark to contextualize our approach against a classical model designed for a similar task. A key takeaway from these results is the parameter efficiency. The quantum model achieves competitive performance with orders of magnitude fewer parameters. However, we acknowledge that with larger datasets or an increased computational budget, the performance landscape could shift; however, for the tasks considered here, the shallow IQP approach offers a remarkably efficient alternative.}

{\section{Quantum device characteristics and demonstration parameters}
\label{app:device_characteristics}

The quantum demonstrations reported in this work were executed on the \texttt{ibm\_aachen} quantum processing unit (QPU), a superconducting quantum computing platform accessed via the IBM Quantum cloud service. In accordance with the guidelines for cloud quantum computing demonstrations, we provide the hardware characteristics at the time of execution. 

\subsection{Device layout and connectivity}
The \texttt{ibm\_aachen} device features a heavy-hexagonal lattice topology. The specific physical layout is illustrated in Fig.~\ref{fig:device_layout}. 

\begin{figure}[htpb]
    \centering
    \includegraphics[width=0.5\linewidth]{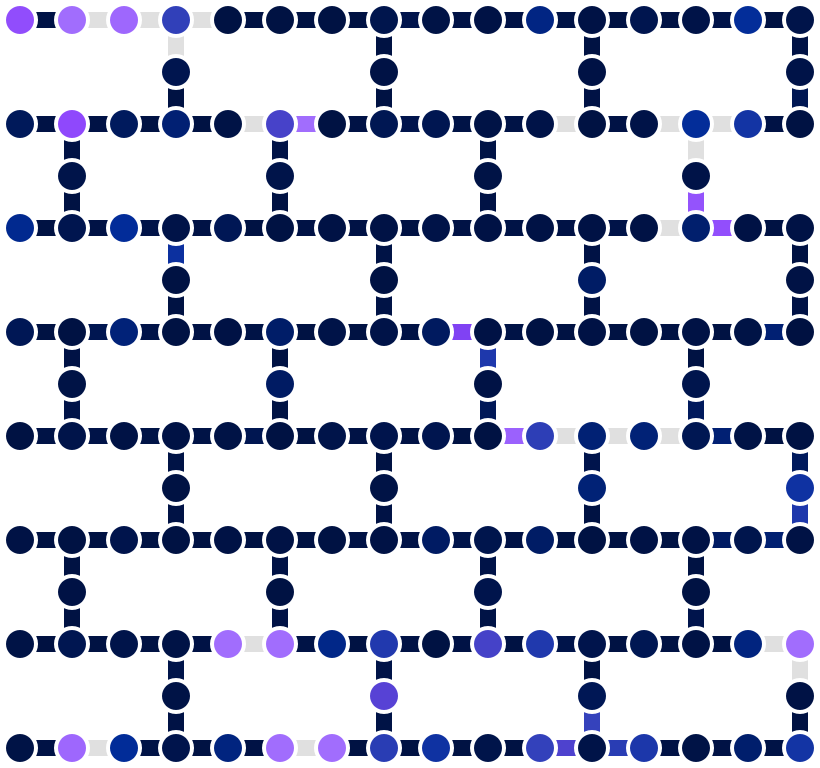}
    \caption{Topology and connectivity graph of the \texttt{ibm\_aachen} QPU at the time of the demonstration. Nodes represent physical qubits, with the color map indicating readout assignment error. Edges denote the RZZ error. In both cases, darker color signifies smaller errors.}
    \label{fig:device_layout}
\end{figure}

\subsection{Coherence and gate fidelities}
Device parameters fluctuate over time; the values presented here represent the state of the QPU during our data acquisition. Table~\ref{tab:qpu_params} summarizes the key performance metrics across the active qubits, extracted from the device calibration data provided by the backend. Specific qubit frequencies were not available in the retrieved calibration payload and are thus omitted.

\begin{table}[htpb]
    \centering
    \caption{Summary of \texttt{ibm\_aachen} device parameters during the demonstration. Values represent the range, median, and mean extracted from the device calibration dataset.}
    \label{tab:qpu_params}
    \begin{tabular}{lccc}
        \hline\hline
        Parameter & Range & Median & Mean \\
        \hline
        Relaxation time, $T_1$ (\textmu s) & $2.46 - 499.57$ & $219.10$ & $220.12$ \\
        Dephasing time, $T_2$ (\textmu s) & $3.35 - 485.63$ & $182.58$ & $189.25$ \\
        Single-qubit gate error & $1.0 \times 10^{-4} - 3.18 \times 10^{-2}$ & $2.0 \times 10^{-4}$ & $7.0 \times 10^{-4}$ \\
        Two-qubit (CZ) gate error & $9.0 \times 10^{-4} - 1.91 \times 10^{-1}$ & $2.0 \times 10^{-3}$ & $1.05 \times 10^{-2}$ \\
        Readout assignment error & $0.0017 - 0.5000$ & $0.0108$ & $0.0681$ \\
        Prob meas0 prep1 & $0.0000 - 1.0000$ & $0.0132$ & $0.0836$ \\
        Prob meas1 prep0 & $0.0000 - 1.0000$ & $0.0045$ & $0.0534$ \\
        Single-qubit gate length (ns) & $32$ & $32$ & $32$ \\
        Readout length (ns) & $2600$ & $2600$ & $2600$ \\
        \hline\hline
    \end{tabular}
\end{table}

\subsection{Data analysis and mitigation}
For this demonstration, the raw bitstrings obtained from the \texttt{ibm\_aachen} QPU were used directly to evaluate the circuits. No algorithmic error mitigation, readout error mitigation, or other post-processing techniques were applied to the results. The data presented reflects the bare hardware performance under the calibrated conditions specified above.}
\end{document}